\documentclass{webofc}
\usepackage[utf8]{inputenc}
\usepackage[varg]{txfonts}
\title{System size and energy dependence of proton rapidity spectra from NA61/SHINE at the CERN SPS}
\usepackage{slantsc}
\usepackage{tikz}

\newcommand{\eV}{\ensuremath{\mbox{e\kern-0.1em V}}\xspace}
\newcommand{\GeV}{\ensuremath{\mbox{Ge\kern-0.1em V}}\xspace}
\newcommand{\MeV}{\ensuremath{\mbox{Me\kern-0.1em V}}\xspace}
\newcommand{\GeVc}{\ensuremath{\mbox{Ge\kern-0.1em V}\!/\!c}\xspace}
\newcommand{\GeVcc}{\ensuremath{\mbox{Ge\kern-0.1em V}\!/\!c^2}\xspace}
\newcommand{\MeVcc}{\ensuremath{\mbox{Me\kern-0.1em V}\!/\!c^2}\xspace}
\newcommand{\AGeV}{\ensuremath{A\,\mbox{Ge\kern-0.1em V}}\xspace}
\newcommand{\AGeVc}{\ensuremath{A\,\mbox{Ge\kern-0.1em V}\!/\!c}\xspace}
\newcommand{\MeVc}{\ensuremath{\mbox{Me\kern-0.1em V}/c}\xspace}

\newcommand{\Geant}{{\scshape Geant}\xspace}

\newcommand{\Epos}{{\scshape Epos}\xspace}

\newcommand{\PHSD}{{\scshape Phsd}\xspace}
\newcommand{\SmashModel}{{\scshape Smash}\xspace}

\newcommand{\CernVM}{\textsc{Cern\-\kern-0.05emVM}\xspace}

\newcommand{\TeV}{\ensuremath{\mbox{Te\kern-0.1em V}}\xspace}

\usepackage{multirow}
\usepackage{lineno}

\usepackage{subcaption}

\author{\firstname{Oleksandra} \lastname{Panova}\inst{1}\fnsep\thanks{\email{oleksandra.panova@cern.ch}} and \firstname{Maciej} \lastname{Lewicki}\inst{2}
         ~for the NA61/SHINE collaboration}

\institute{Jan Kochanowski University of Kielce \and Institute of Nuclear Physics, Polish Academy of Sciences, Cracow }

\begin{document}
\abstract{NA61/SHINE is an experiment at the CERN Super Proton Synchrotron. The main goals of the experiment are the search for the critical point of strongly interacting matter and the study of the properties of the onset of deconfinement. To reach these goals, the two-dimensional scan in beam momentum ($13A-150A$ GeV/$c$) and system size ($p$+$p$, Be+Be, Ar+Sc, Xe+La, Pb+Pb) was performed.

In the final stage of the collision, the spectra of protons are only weakly affected by the effects of resonance decays and rescattering due to their large mass.
Thus, proton rapidity distribution is particularly sensitive to the onset of deconfinement.

This article presents experimental results on proton production in the collision energy range, which is most relevant to the onset of deconfinement. The procedure of measuring the proton rapidity spectra by NA61/SHINE is described, as well as Collaboration's recent results from reactions of $p$+$p$,  Be+Be and Ar+Sc. Presented experimental results are confronted with existing data and models.}
\maketitle

\section{Introduction}
\label{sec:introduction}

NA61/SHINE is a fixed target experiment \cite{NA61, NA61_1} located in CERN’s SPS North Area. It is a multi-purpose spectrometer measuring particle production in hadron-hadron, hadron-nucleus, and nucleus-nucleus collisions in the energy range $\sqrt{s_{NN}}=5.12$--$17.3~\mathrm{GeV}$. The goals of the experiment include the search for the critical
point of strongly interacting matter and the study of properties of the onset of
deconfinement. To reach these goals, the two-dimensional scan in collision energy and system size 
was performed.

The program is mainly motivated by the observation of rapid changes of hadron production properties in central Pb+Pb collisions at about $30A$ GeV/$c$ by the NA49 experiment \cite{NA49}: a sharp peak in the $K^+/\pi^+$ ratio (``horn''), the start of a plateau in the inverse slope parameter for charged kaons (``step''), and a steepening of the increase of pion production per wounded nucleon with increasing collision energy (``kink''), which can be interpreted as the signatures of the onset of deconfinement.  This article for the first time reviews the rich data from NA61/SHINE in the context of yet another possible signature of the onset of deconfinement, which is the evolution of the shape of the proton rapidity spectrum with increasing collision energy.

The article is organized as follows.  Section \ref{sec:protons} discusses the motivation to study the shape of proton rapidity spectra.
In Sec.~\ref{sec:na61}, the procedure of measurement of the proton rapidity spectra is described. The experimental results obtained by NA61/SHINE are presented in Sec.~\ref{sec:na61res}, along with a comparison with available world data. Preliminary results on proton spectra measured in Ar+Sc collisions are shown for the first time. A confrontation with current phenomenological models is discussed in Sec. \ref{sec:models}. A summary in the Sec. \ref{sec:conclusions} closes the article.

\section{Proton rapidity spectra and the onset of deconfinement}
\label{sec:protons}
At SPS energies protons are relatively abundant among products of nuclear collisions and relatively easy to identify (mass is significantly larger than pion and
kaon masses). Their rapidity distributions are weakly affected by processes at the final
stages of collisions, and they were suggested to be sensitive to the onset of deconfinement~\cite{IvanovPLB, Ivanov2}.

Assuming the equation of state that features the phase transition, one expects a particular evolution of the proton rapidity spectra with increasing collision energy ~\cite{Ivanov2}:
\begin{enumerate}[(i)]
    \item At low collision energies, the softest point of the equation of state is not reached, the system is stiff and the produced fireball is almost spherical. Therefore, the proton rapidity distribution has a peak at the midrapidity (Fig.~\ref{fig:sketch1}).
    \item With the increase of collision energy, the equation of state reaches its softest point, the fireball is deformed to the disk shape, the rapidity distribution has two peaks and a dip between them (Fig.~\ref{fig:sketch2}).
    \item Then, with a further increase of collision energy, the equation of state becomes stiff again and the fireball becomes less deformed, more spherical. Rapidity distribution has one peak or, at least, the dip strongly decreases (Fig.~\ref{fig:sketch3}).
    \item Finally, with further increase of energy, kinetic pressure overcomes stiffness of the equation of state -- the fireball is strongly deformed and rapidity distribution has two peaks and a dip between them, again (Fig.~\ref{fig:sketch4}). 
\end{enumerate}
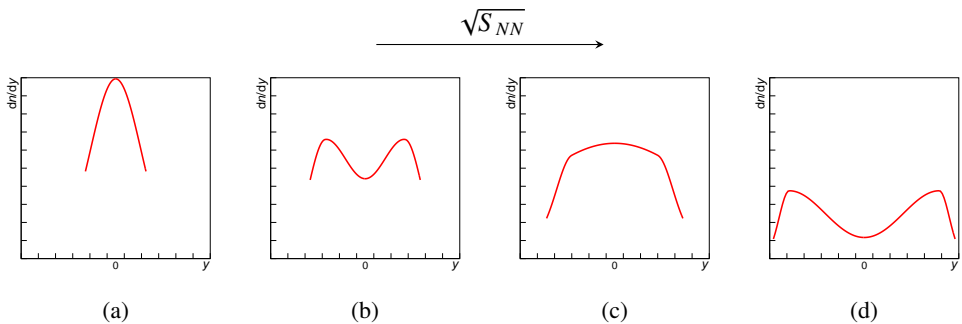
\begin{figure}[h]
     \centering
      \begin{tikzpicture}
            \draw[-stealth] (0,0) -- node[yshift=3mm] {$\sqrt{S_{NN}}$} (3,0) ;
    \end{tikzpicture}\\[1mm]
     \begin{subfigure}[b]{0.24\textwidth}
         \centering
         \includegraphics[width=\textwidth]{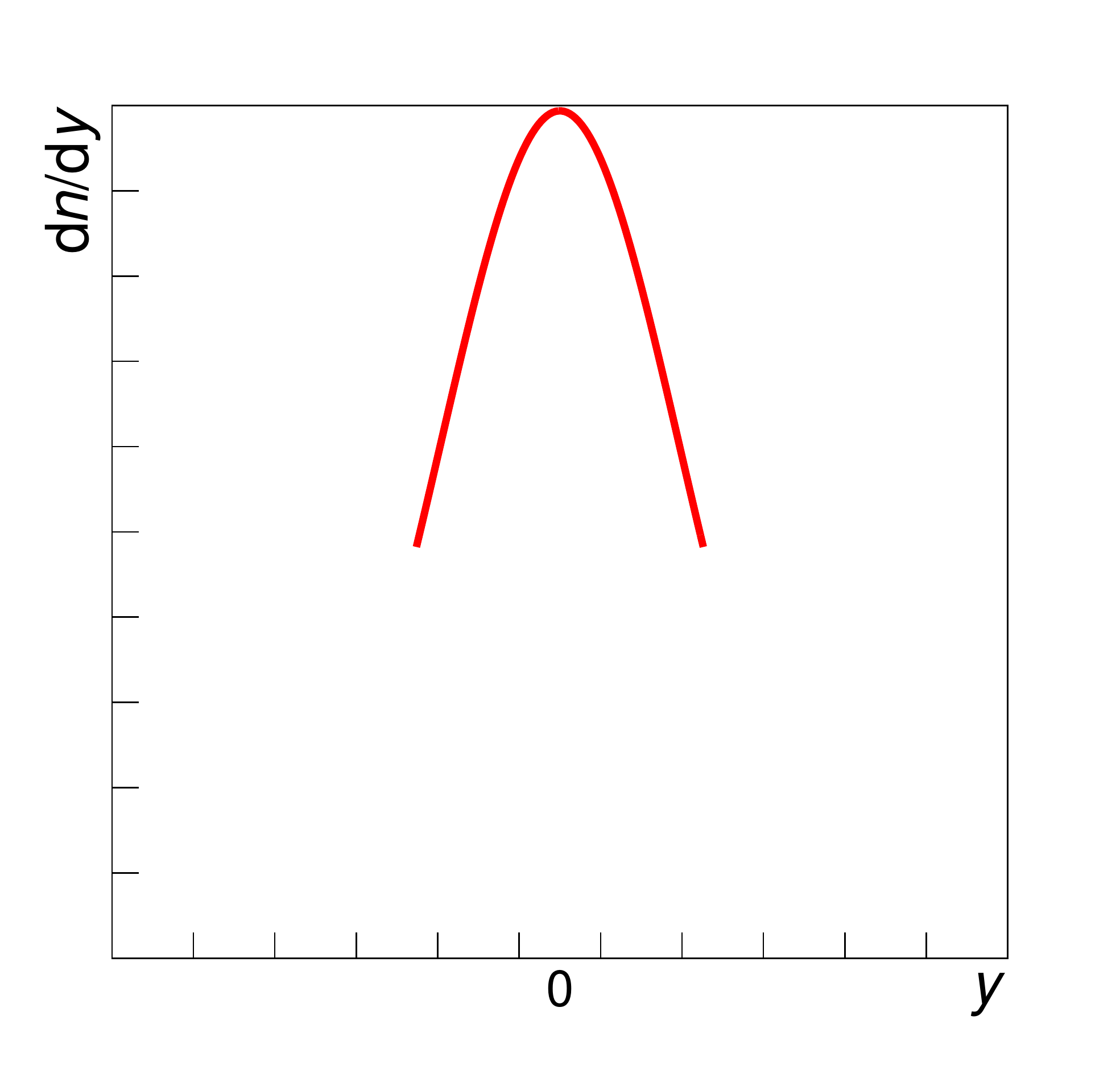}
         \caption{}
         \label{fig:sketch1}
     \end{subfigure}
     \hfill
     \begin{subfigure}[b]{0.24\textwidth}
         \centering
         \includegraphics[width=\textwidth]{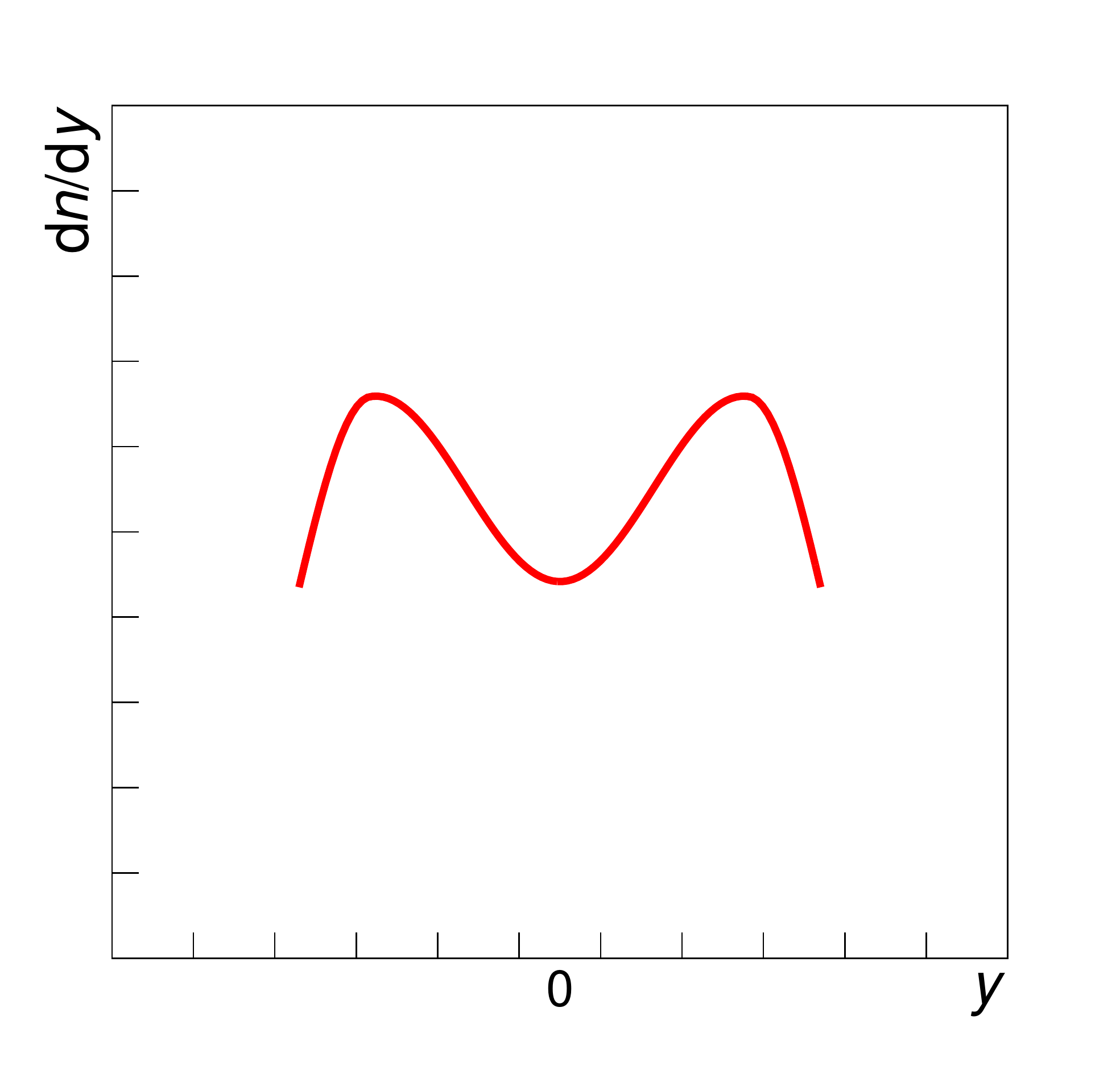}
         \caption{}
         \label{fig:sketch2}
     \end{subfigure}
     \hfill
     \begin{subfigure}[b]{0.24\textwidth}
         \centering
         \includegraphics[width=\textwidth]{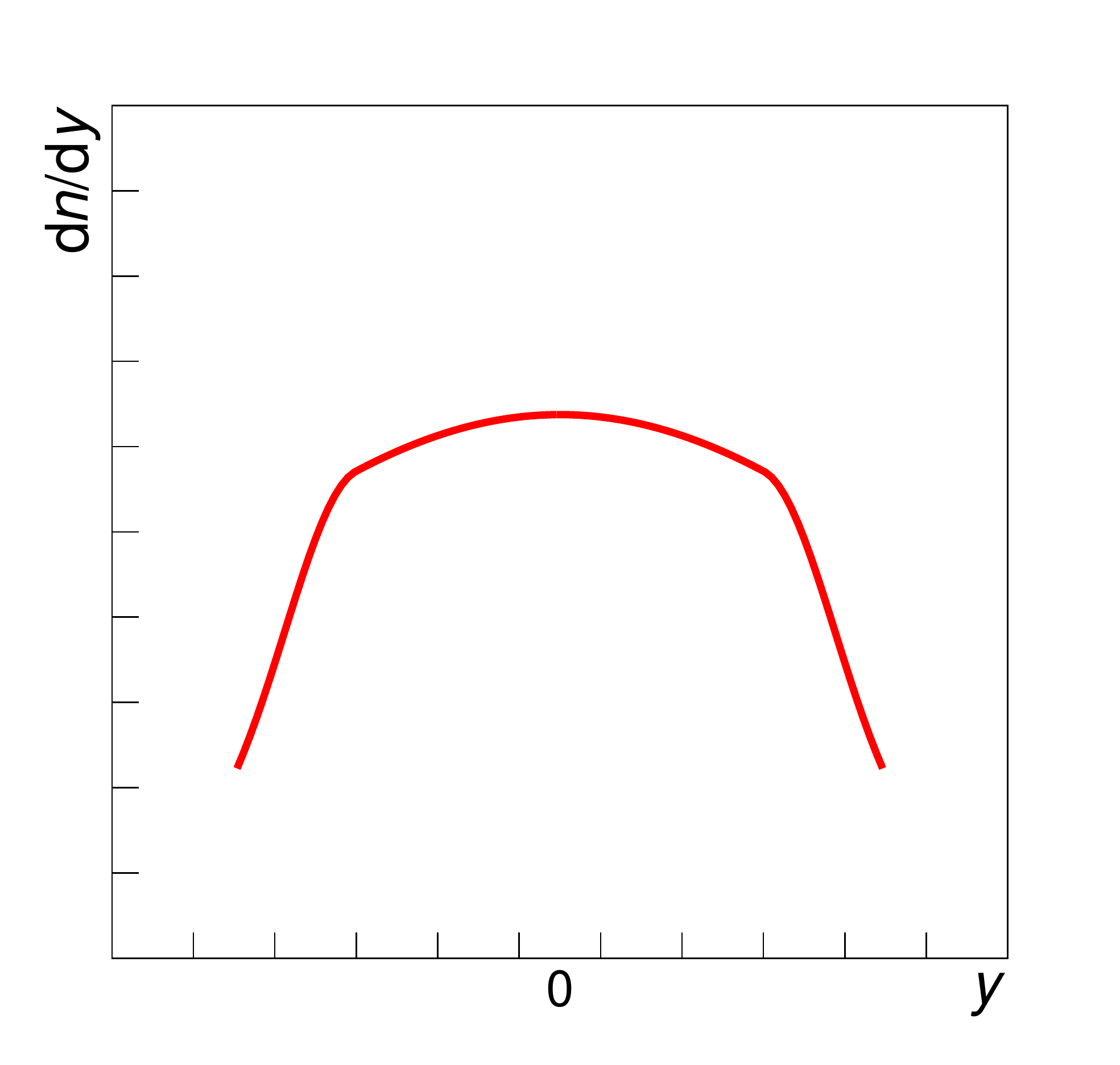}
         \caption{}
         \label{fig:sketch3}
     \end{subfigure}
     \hfill
     \begin{subfigure}[b]{0.24\textwidth}
         \centering
         \includegraphics[width=\textwidth]{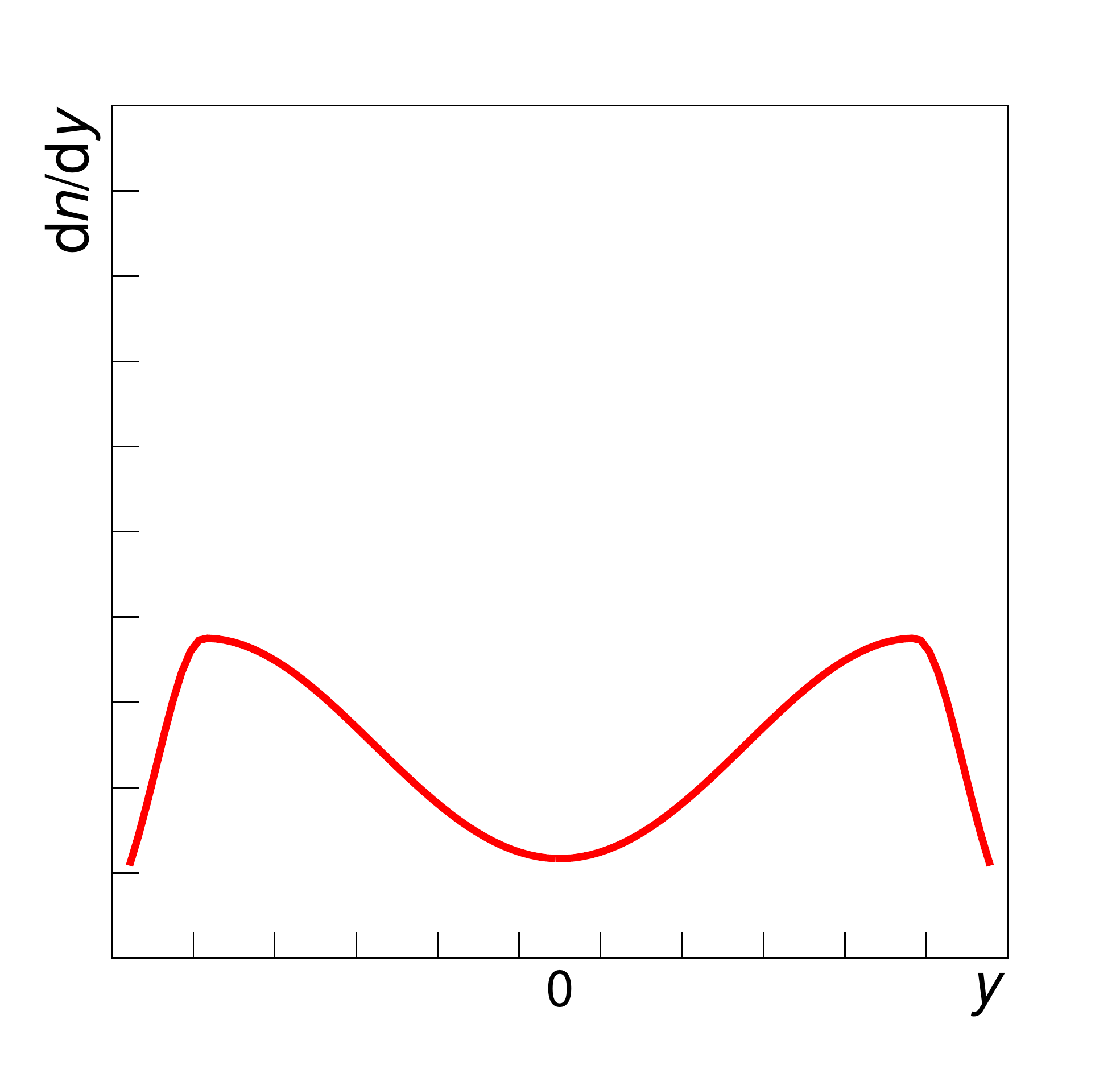}
         \caption{}
         \label{fig:sketch4}
     \end{subfigure}
     
        \caption{Sketch describing ``peak-dip-peak-dip'' irregularity, proposed as a signal of deconfinement in Ref. \cite{Ivanov2}.}
        \label{fig:sketch}
\end{figure}
This behaviour of spectra (called here ``peak-dip-peak-dip'' irregularity) exists only for the two-phase equation of state, thus it can be viewed as one of the signatures of the onset of deconfinement.

\section{Measurements of proton rapidity spectra in NA61/SHINE}
\label{sec:na61}

The geometry of the NA61/SHINE detector allows measuring a large fraction of particles produced into
the forward hemisphere of the collision, down to zero transverse momentum
$p_T=0~\mathrm{GeV}/c$, allowing for the calculation of mean multiplicities of hadrons
produced in the whole phase-space (4$\pi$).
The main tracking components of the detector are four Time Projection Chambers (TPC), two of which are located in the magnetic field. Additional measurements for particle identification are done using Time of Flight (ToF) detectors. The Projectile Spectator Detector (PSD), a high-resolution forward zero-degree calorimeter, is used to determine the centrality of the collisions.

Charged particle identification in the NA61/SHINE experiment is primarily based on the measurement of the ionization energy loss, $\mathrm{d}E/\mathrm{d}x$, in the region of momentum above  $5~\mathrm{GeV}/c$. At lower momenta, the $\mathrm{d}E/\mathrm{d}x$ bands for different particle species overlap and additional measurement of time of flight (\textit{tof}) obtained from the ToF walls is required to remove the ambiguity. These two methods allow covering most of the phase space in rapidity and transverse momentum which is of interest for the strong interaction program of NA61/SHINE. 

Only central events with well-positioned beam without off-time particles were chosen for the analysis, with an additional requirement on the quality and position of primary vertex fit. Furthermore, a number of track selection criteria were applied to reduce the background from  non-primary particles and to maximize the tracking efficiency and the $\mathrm{d}E/\mathrm{d}x$ resolution.

The raw experimental results were corrected using the Monte Carlo (MC) simulation to account for the effects of detector geometry, inefficiency and contribution from non-primary particles. First, inelastic interactions were generated using \Epos model \cite{EPOS}, followed by the \Geant3-based \cite{geant} program chain used to track particles through the spectrometer, generate decays and secondary interactions and finally the detector response was simulated. Then, the simulated events were processed using the standard NA61/SHINE reconstruction chain and reconstructed tracks were matched to the simulated particles based on the cluster positions. The simulated dataset is used to derive the corrections to the raw data. The heavily model-dependent contribution from weak decay feed-down was studied carefully and adjusted using data-driven methods. The dependence on the model of the corrections due to detector-related effects is minimal.

Total uncertainties were calculated as a combination of statistical and systematic uncertainties. For systematic uncertainties following sources were considered: particle identification method, event and track selection, and feed-down corrections.

\section{NA61/SHINE experimental results}
\label{sec:na61res}

Currently, NA61/SHINE measured proton rapidity spectra for three systems:
\begin{enumerate}[(i)]
    \item inelastic $p$+$p$ collisions at $19A-158A$ GeV/$c$ published in Refs.~\cite{pp1, pp2},
    \item central (20\%) Be+Be collisions at $19A-150A$ GeV/$c$ published in Refs.~\cite{bebe1,bebe2},
    \item central (10\%) Ar+Sc  collisions at $13A-150A$ GeV/$c$ preliminary results shown here for the first time.
\end{enumerate}

\begin{figure}[h!]
    \centering
    \includegraphics[width = 0.327\linewidth]{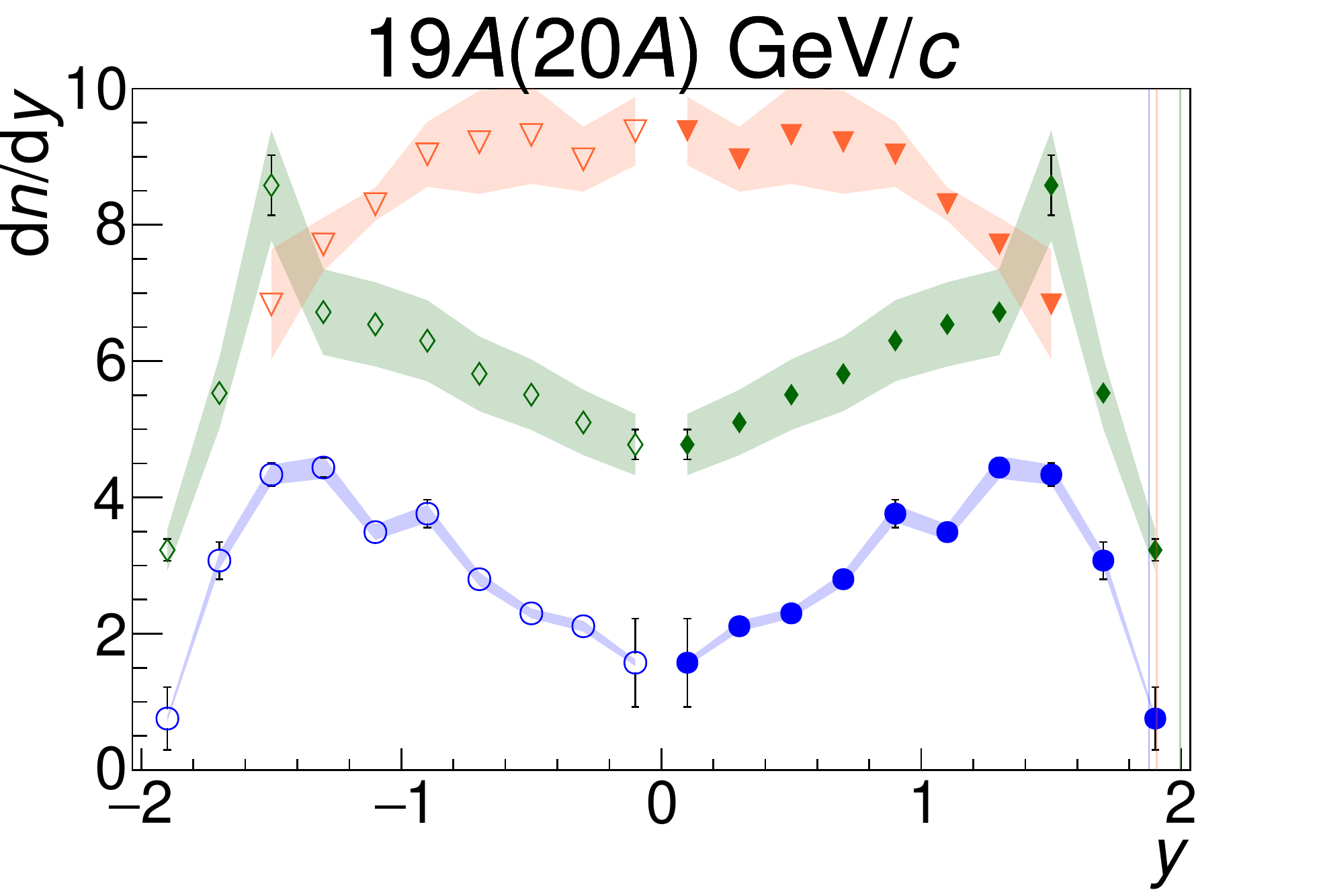}
    \includegraphics[width = 0.327\linewidth]{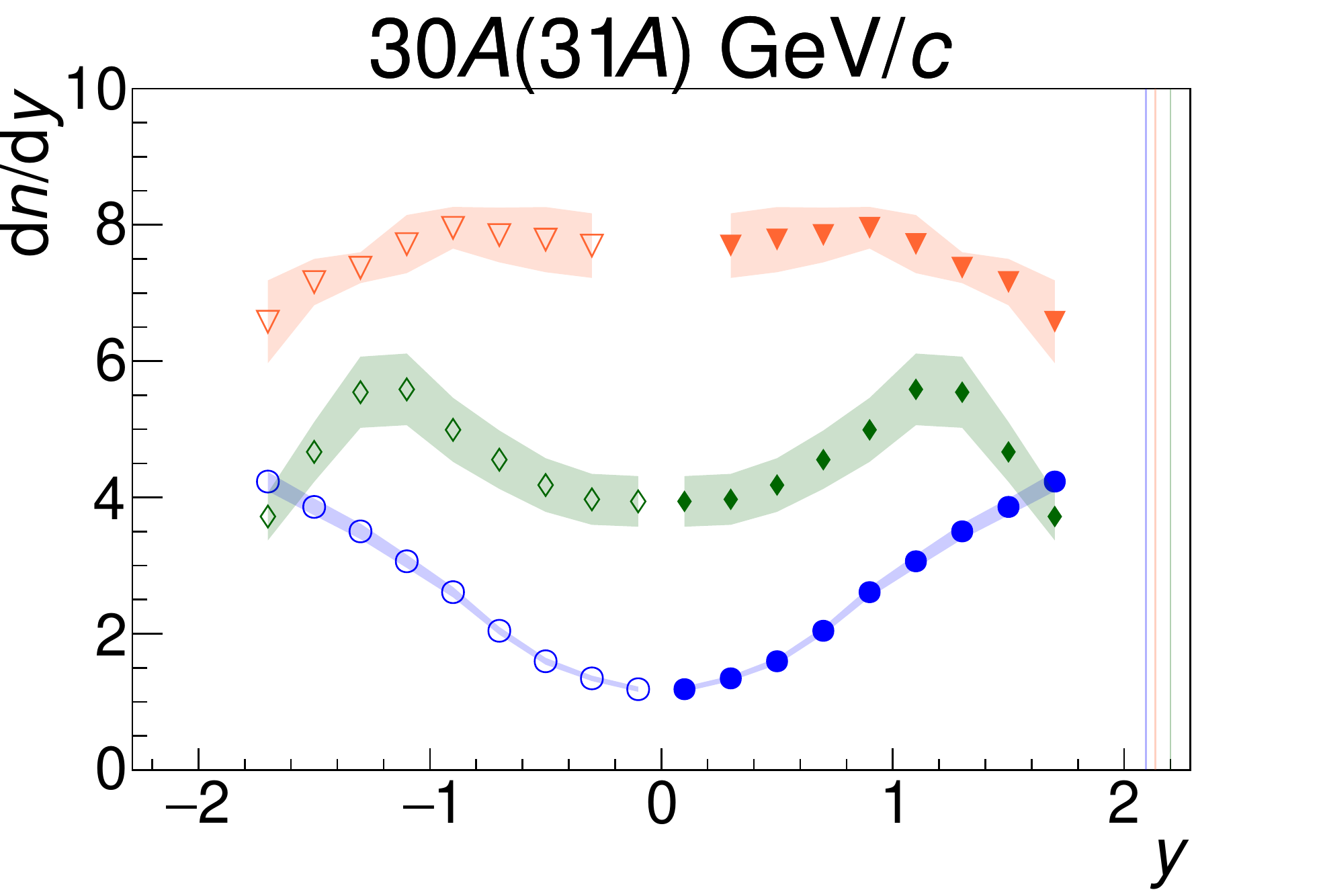}
    \includegraphics[width = 0.327\linewidth]{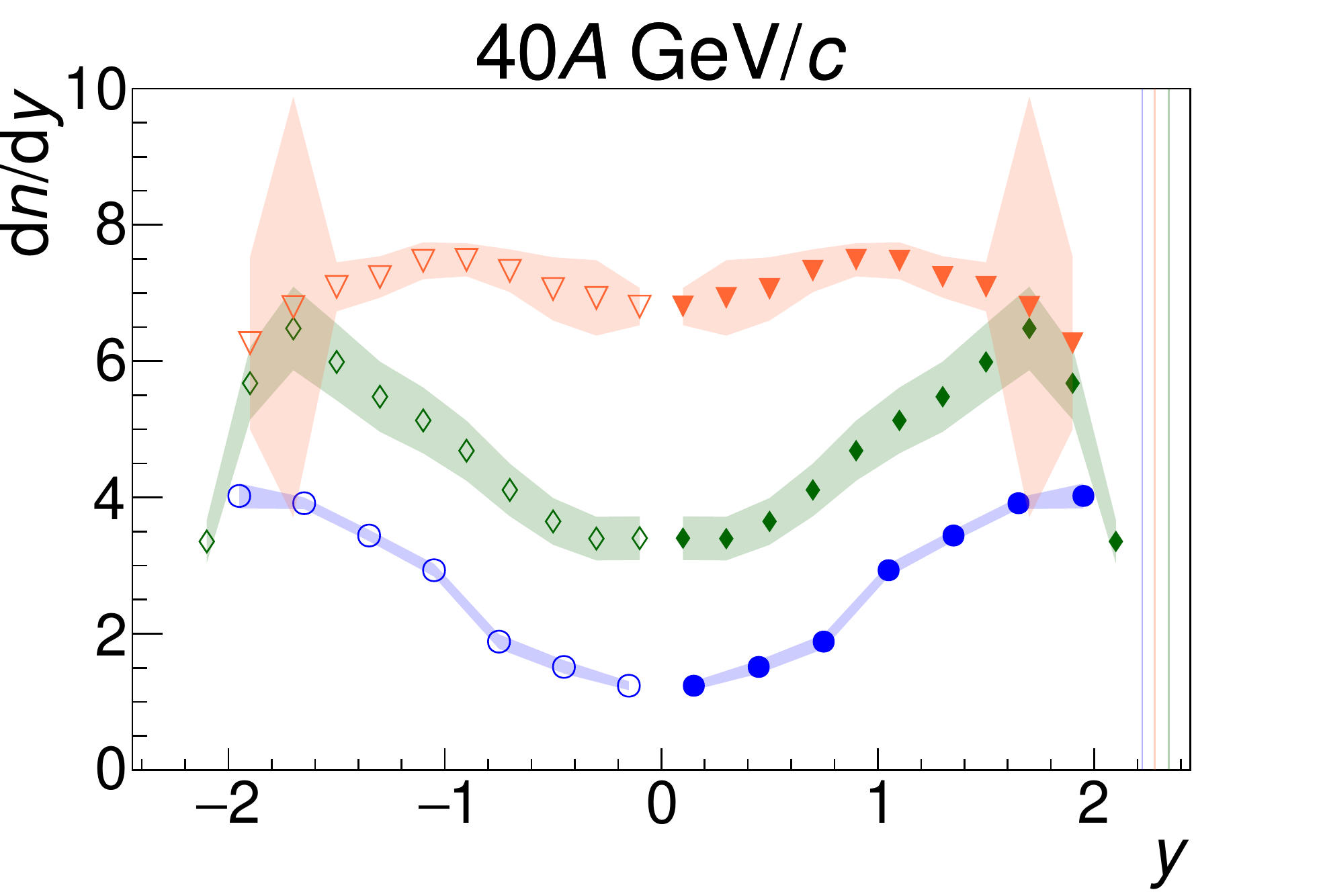}\\
    \hspace{-0.5cm}
    
    \includegraphics[width = 0.15\linewidth,trim={2.8cm 7.8cm 2.8cm 3cm}, clip]{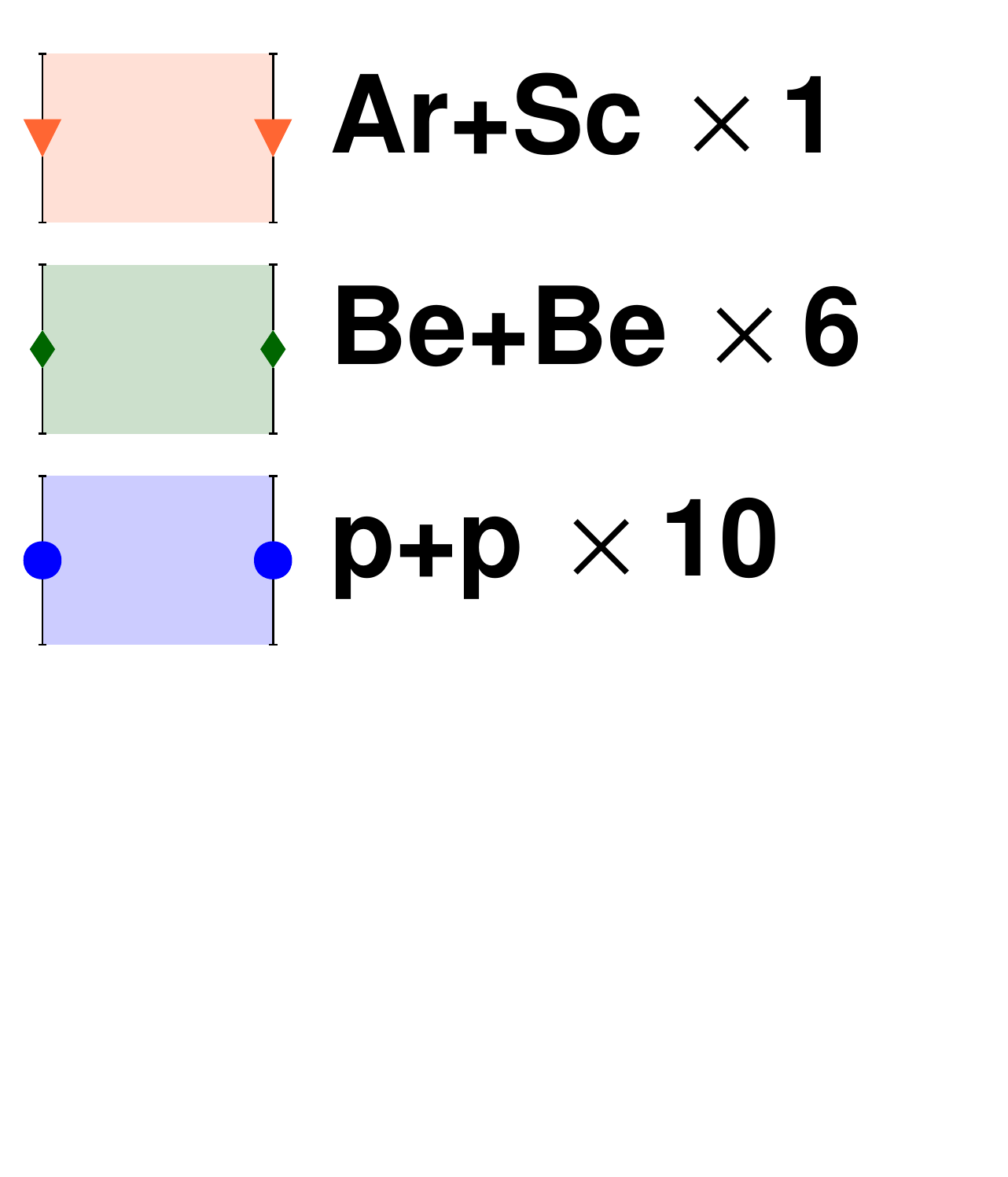} 
    \includegraphics[width = 0.327\linewidth]{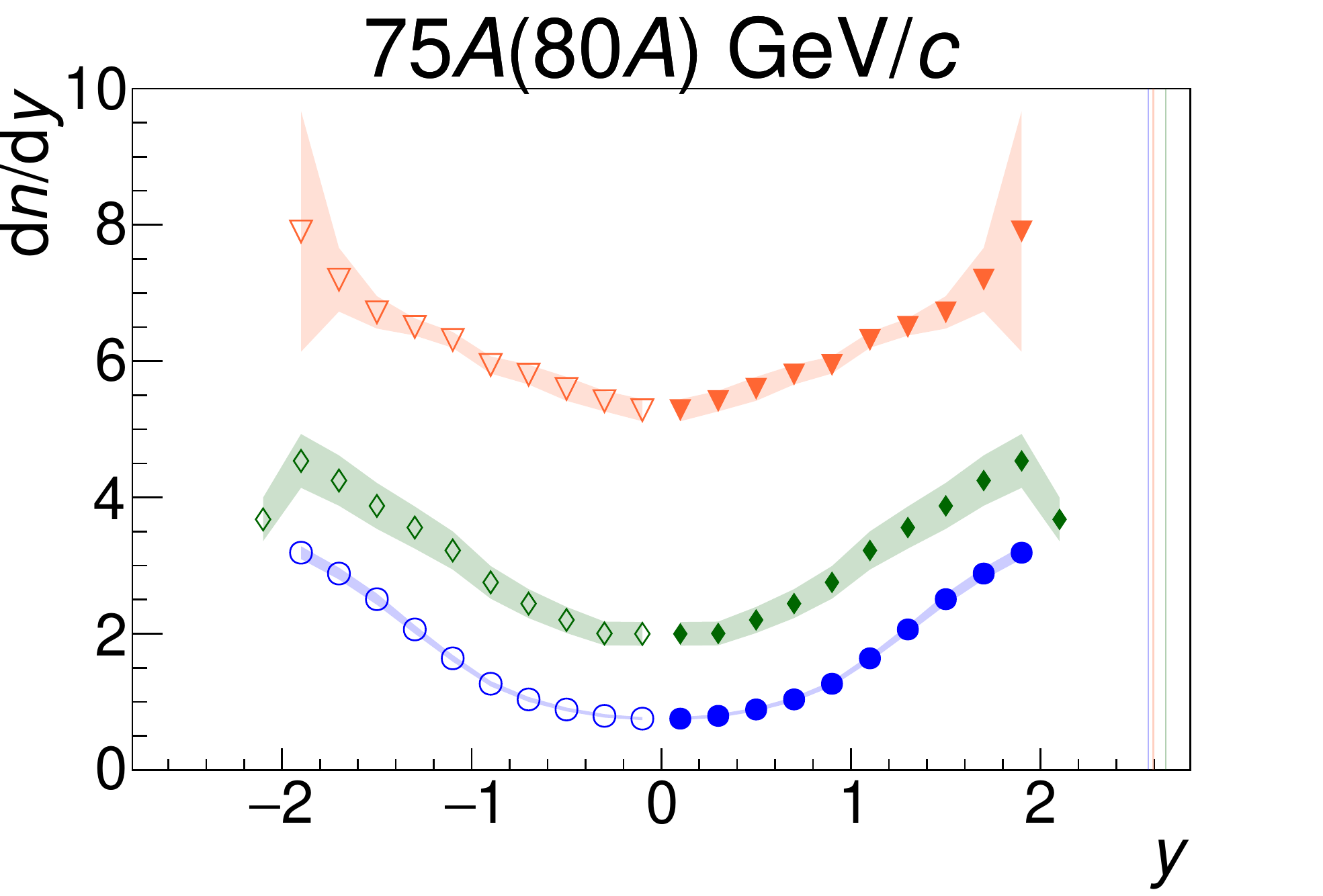}
    \includegraphics[width = 0.327\linewidth]{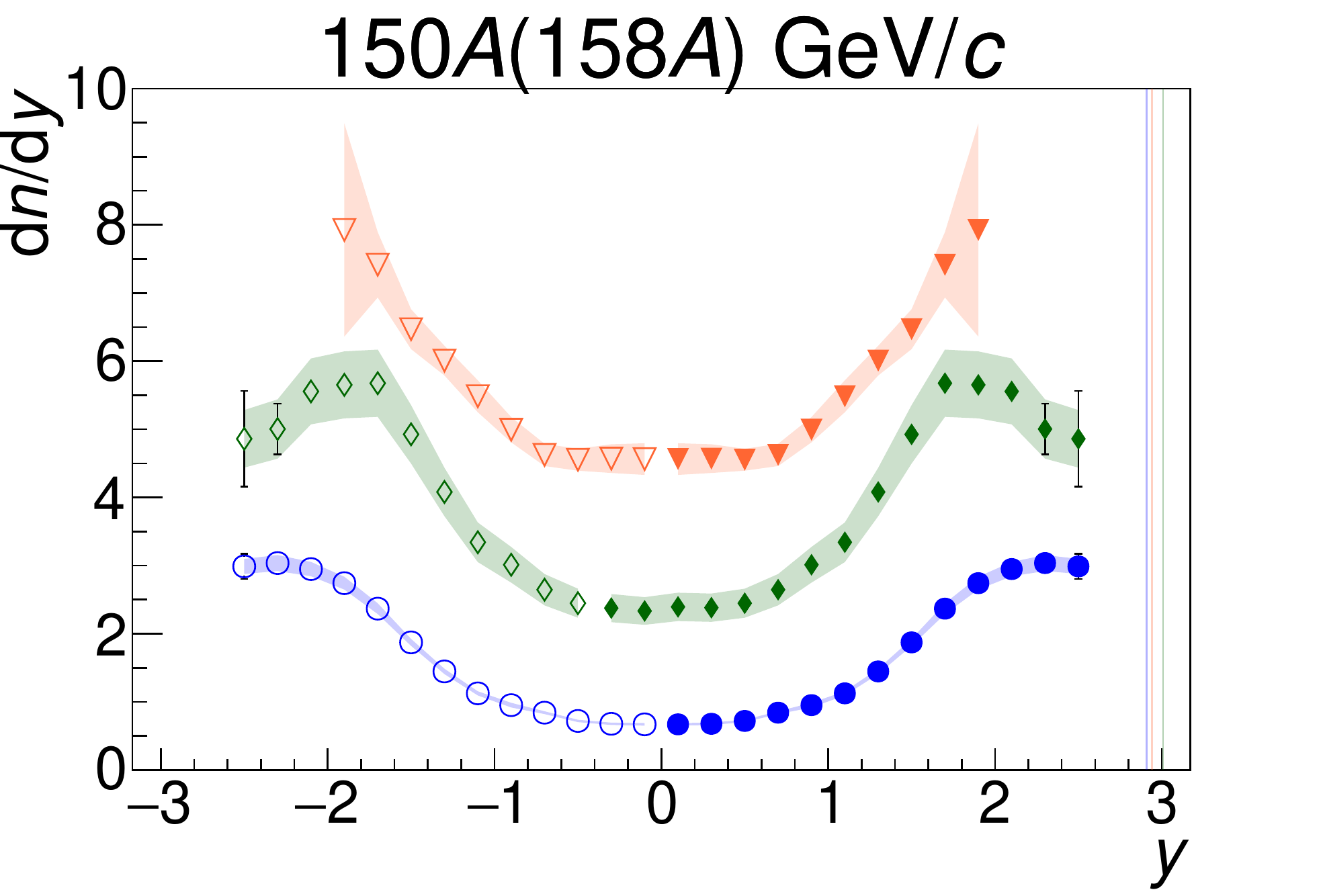}\hspace{3cm}
    \caption{Rapidity spectra of protons produced in inelastic $p$+$p$ and central Be+Be and Ar+Sc (preliminary results) collisions at beam momenta in range $19A(20A)-150A(158A)$~GeV/$c$. See Sec. \ref{sec:na61res} for more details. 
    }
    \label{proton_y}
\end{figure}

Figure \ref{proton_y} presents rapidity spectra of protons produced in inelastic  $p$+$p$ and central Be+Be and Ar+Sc collisions at beam momenta $19A(20A),~30A(31A),~40A,~75A(80A),$ $~150A(158A)$~GeV/$c$. Full points are
measured data, open points show a reflection with respect to mid-rapidity. Systematic and statistical uncertainties are shown as shaded bands and bars, respectively. Vertical coloured lines are plotted on the beam rapidity of the corresponding system.

\begin{figure}[h!]
    \centering
    \includegraphics[width = 0.495\linewidth]{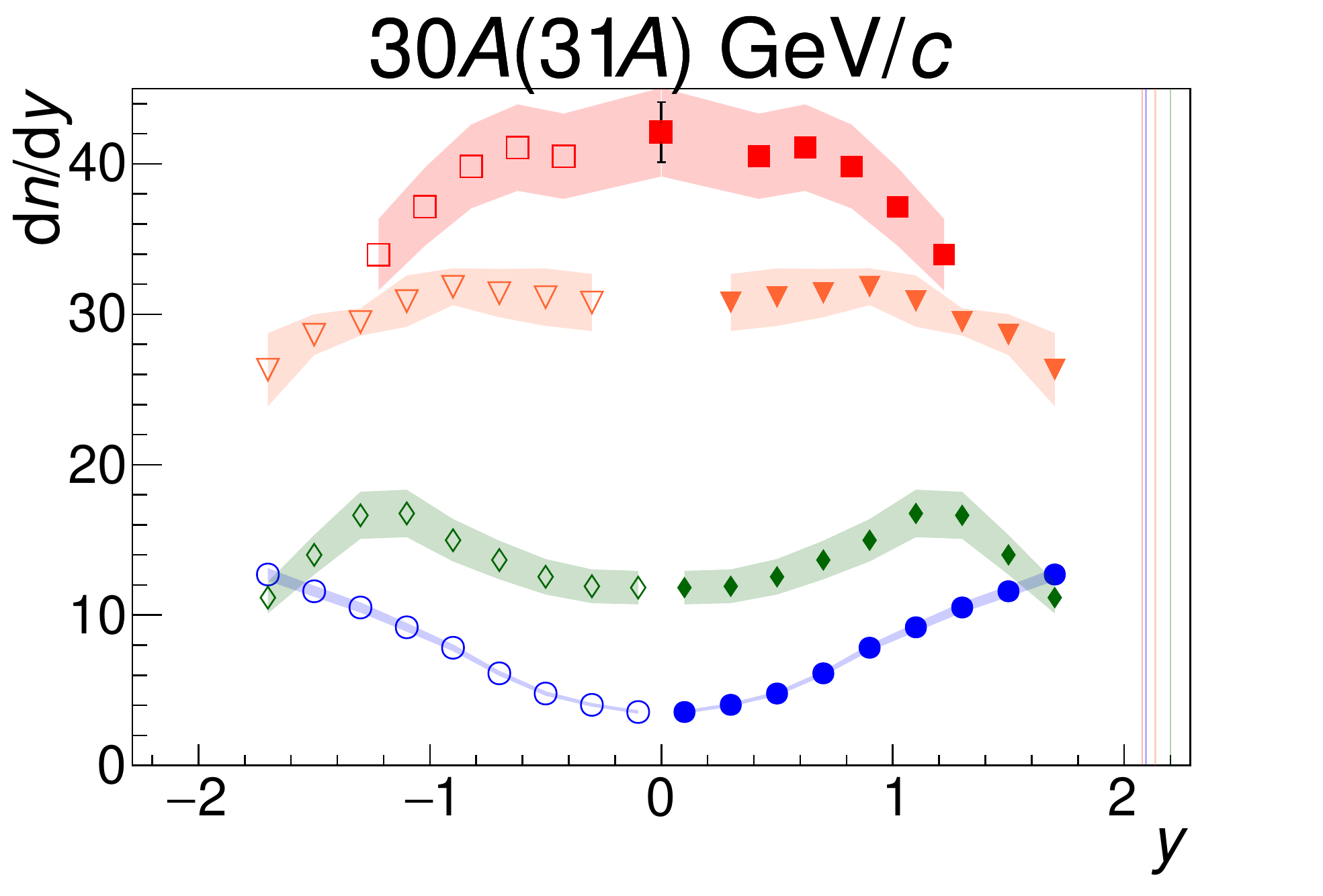}
    \includegraphics[width = 0.495\linewidth]{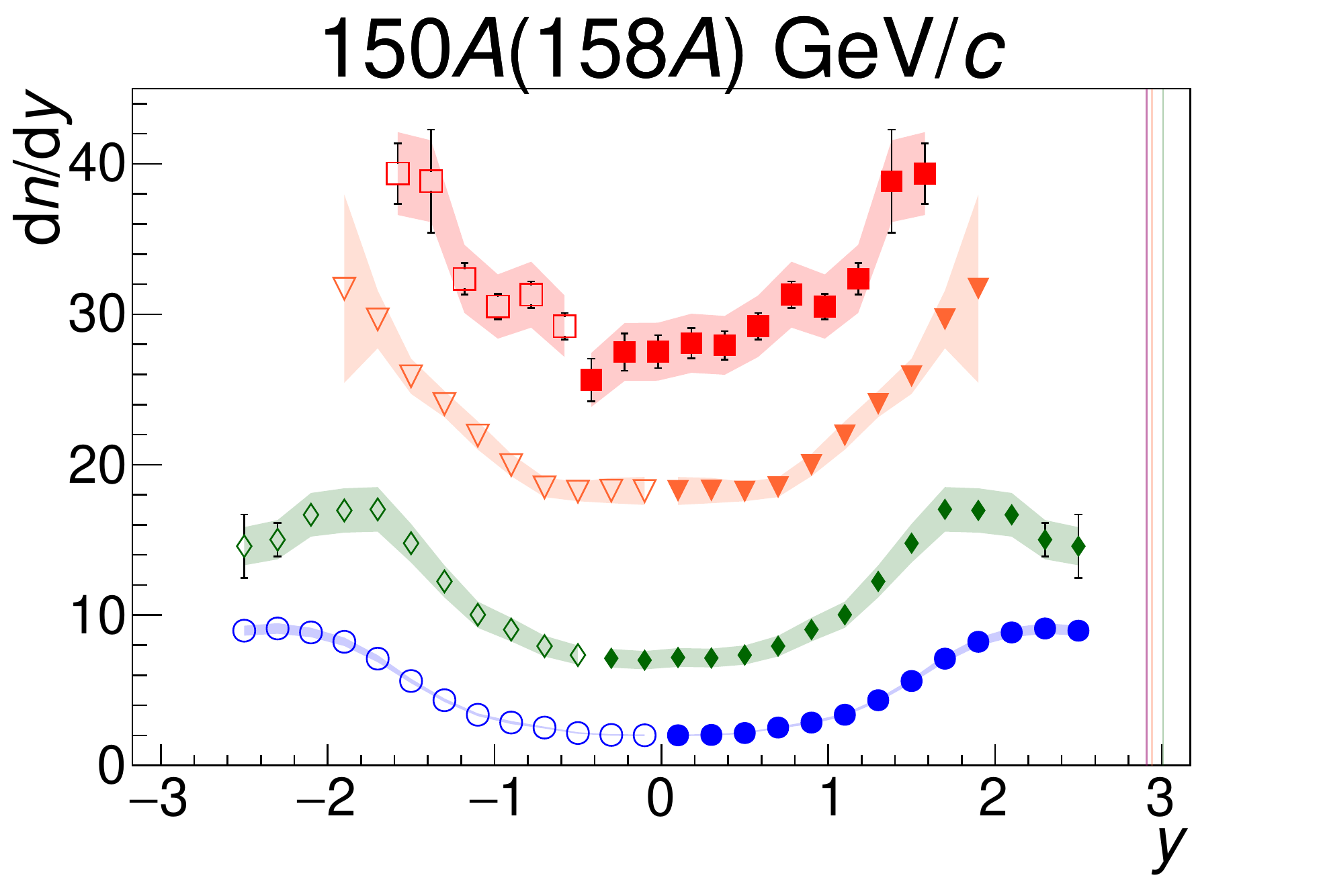}\\
    \includegraphics[width = 0.5\linewidth,trim={2.3cm 0.39cm 2.199cm 0.35cm}, clip]{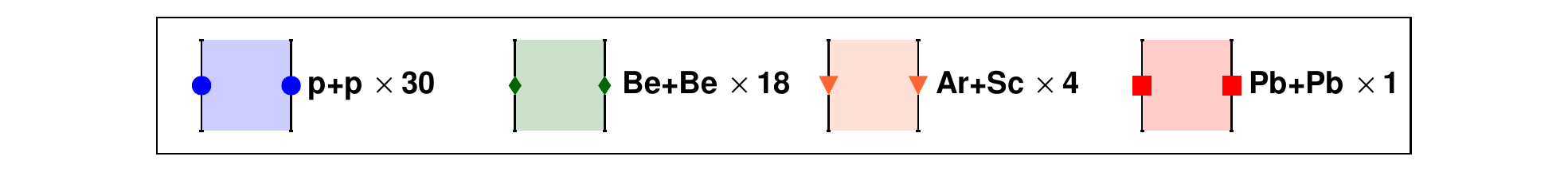}
    \caption{Rapidity spectra of protons produced in inelastic $p$+$p$ and central Be+Be, Ar+Sc (preliminary) and Pb+Pb (preliminary for 30$A$ GeV/$c$) collisions at 30$A$(31$A$) GeV/$c$ (\textit{left}) and 150$A$(158$A$) GeV/$c$ (\textit{right}).}
    \label{fig:pb}
\end{figure}

Figure \ref{fig:pb} presents a comparison of proton rapidity spectra measured by NA61/SHINE with results on central (5\%) Pb+Pb collisions measured by NA49 at two beam momenta, $30A$ GeV/$c$ (preliminary) and $158A$ GeV/$c$ \cite{na49_proton}.

In the collisions of light systems, $p$+$p$ and Be+Be, the proton rapidity spectra feature a dip at midrapidity for all energies, while in the case of Ar+Sc collisions, the transition from the ``peak'' to the ``dip'' shape is clearly visible.
Qualitatively, the Ar+Sc spectra are evidently more similar to the ones from Pb+Pb collisions, than to spectra from Be+Be interactions. For both, Ar+Sc and Pb+Pb reactions, the ``peak-dip'' transition is observed within the SPS energy range. 
However, the ``peak-dip-peak-dip'' irregularity is observed for the heaviest systems, as the data is available at a wider collision energy range (Pb+Pb at SPS \cite{na49_proton} and Au+Au at AGS \cite{ags}), what probably might also be the case of Ar+Sc as well (within the beam energy range of future CBM facility at FAIR). 

\section{System size and collision energy dependence}

\begin{figure}[h!]
    \centering
    \includegraphics[width = 0.425\linewidth,trim={0cm 0cm 1.399cm 0cm}, clip]{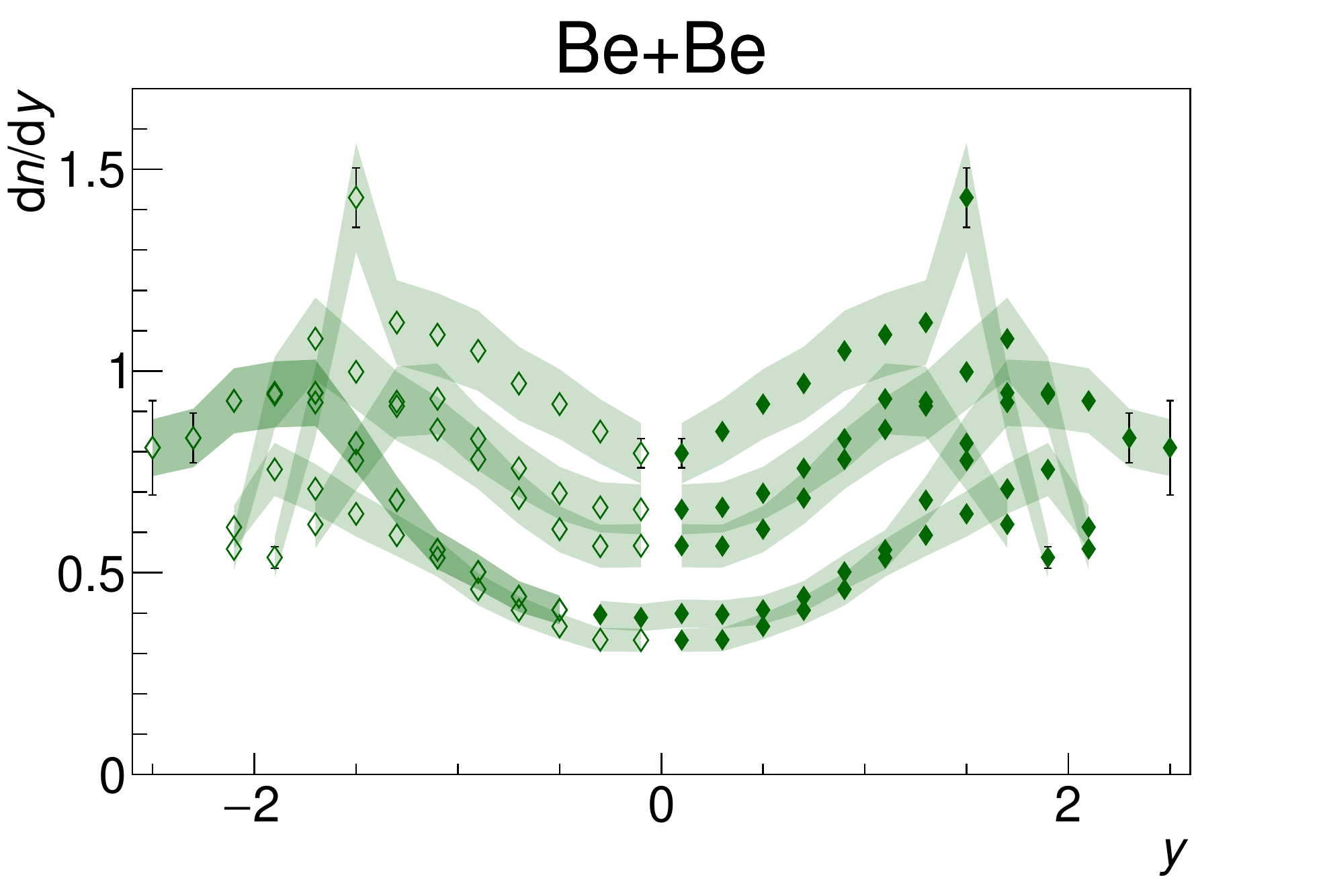}
    \hspace{-0.44cm}
    \begin{tikzpicture}[scale=1.15]
        \path[->,thick] (0,1.6) edge  [] (0,-1);
        \node[below] at (0.27,-0.96) {\scriptsize $p$, GeV/$c$};
        \node[right] at (0,1.0) {\scriptsize $19A$};
        \filldraw (0,1.0) circle (1pt);
        \node[right] at (0,-0.4) {\scriptsize $150A$};
        \filldraw (0,-0.4) circle (1pt);
        \filldraw (0,-1.5) circle (0pt);
    \end{tikzpicture}
    \includegraphics[width = 0.425\linewidth,trim={0cm 0cm 1.399cm 0cm}, clip]{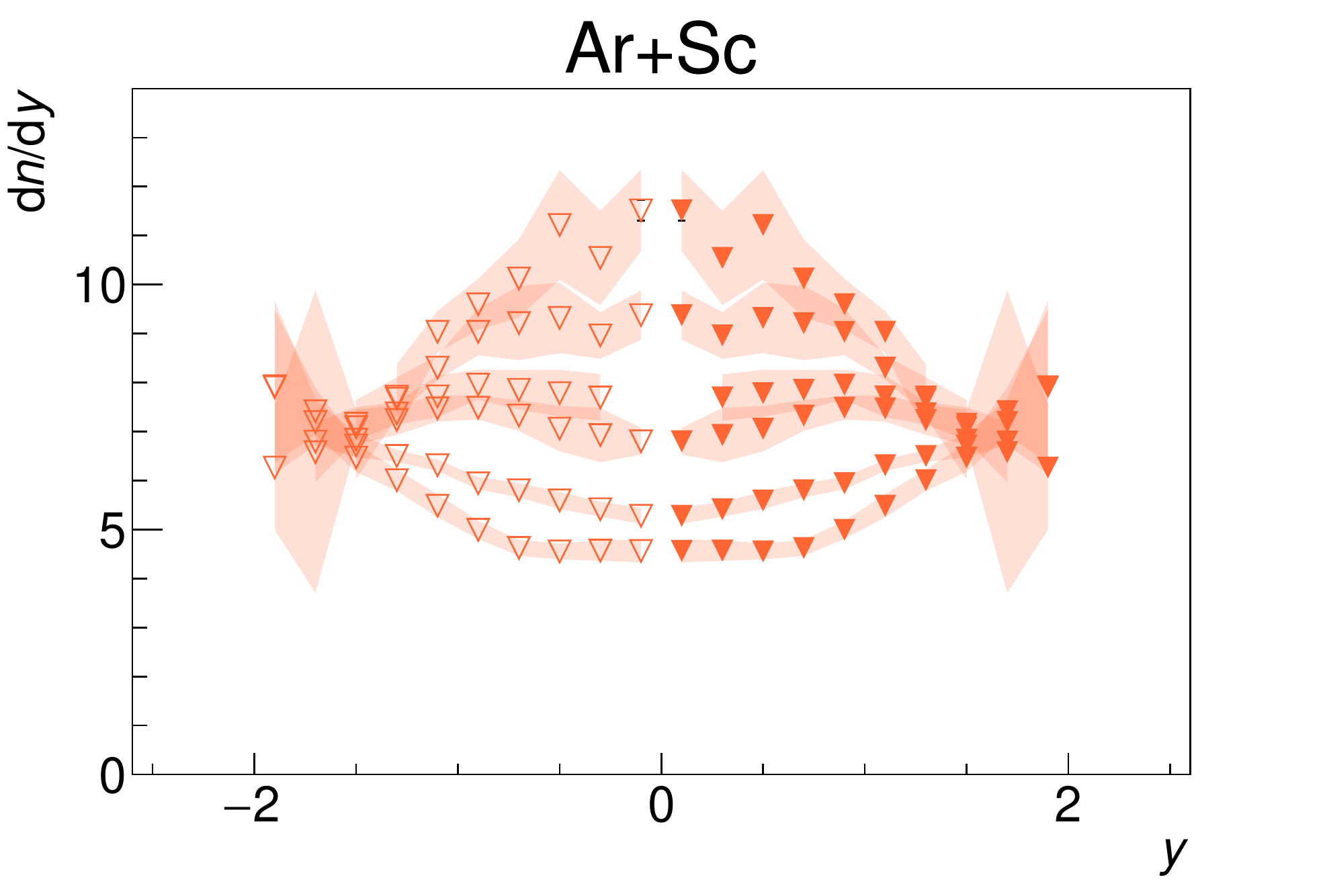}
    \hspace{-0.44cm}
    \begin{tikzpicture}[scale=1.15]
        \path[->,thick] (0,1.6) edge  [] (0,-1);
        \node[below] at (0.27,-0.96) {\scriptsize $p$, GeV/$c$};
        \node[right] at (0,1.05) {\scriptsize $13A$};
        \filldraw (0,1.05) circle (1pt);
        \node[right] at (0,-0.1) {\scriptsize $150A$};
        \filldraw (0,-0.1) circle (1pt);
        \filldraw (0,-1.5) circle (0pt);
    \end{tikzpicture}\\[-41.5mm]
    ~\hspace{0.65\linewidth} {\color{gray}\scriptsize (preliminary)}\\[37mm]
    \caption{Two qualitatively different trends in energy dependence: no ``peak-dip'' transition for Be+Be (\textit{left}) and ``peak-dip'' transition for Ar+Sc (\textit{right}) collisions.}
    \label{fig:2cl}
\end{figure}
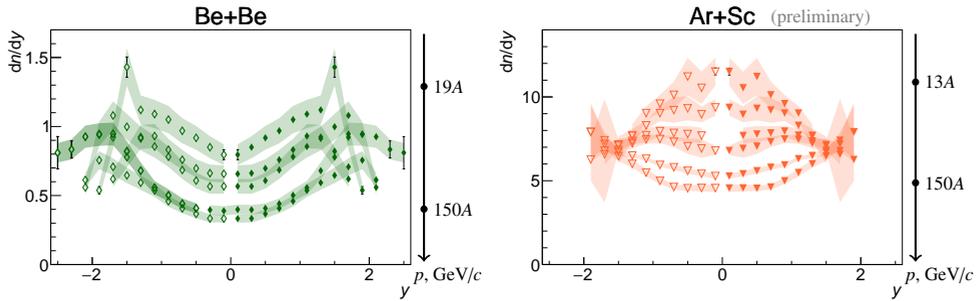

The system size dependence observed in the results from NA49 and NA61/SHINE experiments can be summarized as follows:
\begin{enumerate}[(i)]
    \item There is no ``peak-dip'' transition for small systems ($p$+$p$ and Be+Be). 
    \item There is a ``peak-dip'' transition for medium and heavy systems (Ar+Sc and Pb+Pb).
\end{enumerate}
A clear visualization of this observation is presented in Fig.~\ref{fig:2cl}.
\label{sec:dep}

Table \ref{tab} summarizes data on the shape of the rapidity distributions of protons produced in inelastic $p$+$p$ and central Be+Be and Ar+Sc collisions measured by NA61/SHINE and central Pb+Pb collisions within the SPS energy range measured by NA49.

There are clear qualitative differences between small ($p$+$p$, Be+Be) and large (Ar+Sc, Pb+Pb) systems -- while the shape of small systems does not change within the SPS collision energy range, the spectra from heavier systems feature a ``peak-dip'' transition.
\newcommand{\wdth}{1.5cm}
    \begin{table}[h!]
    \centering
    \begin{tabular}{|r|r|c|c|c|c|}
    \hline
    \multicolumn{2}{|c|}{}&\multicolumn{3}{c|}{\textbf{NA61/SHINE}} & NA49\\
    \hline
     $p_{Lab},$     &  $\sqrt{s_{NN}},$           & \multirow{2}{*}{$p$+$p$} & \multirow{2}{*}{Be+Be} & Ar+Sc & Pb+Pb \\ 
      GeV/$c$     &   GeV          &  &  &{\scriptsize(preliminary)}&{\scriptsize(20$A$, 31$A$ preliminary)}  \\ \hline
    20$A$ (19$A$)  &  6.3        &  \includegraphics[width=\wdth]{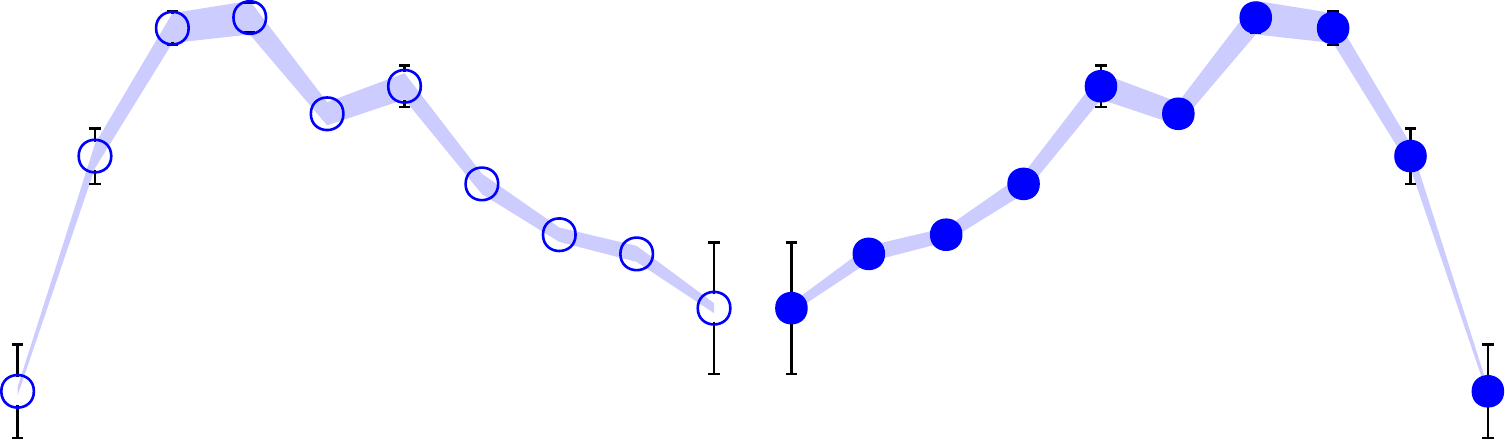} &  \includegraphics[width=\wdth]{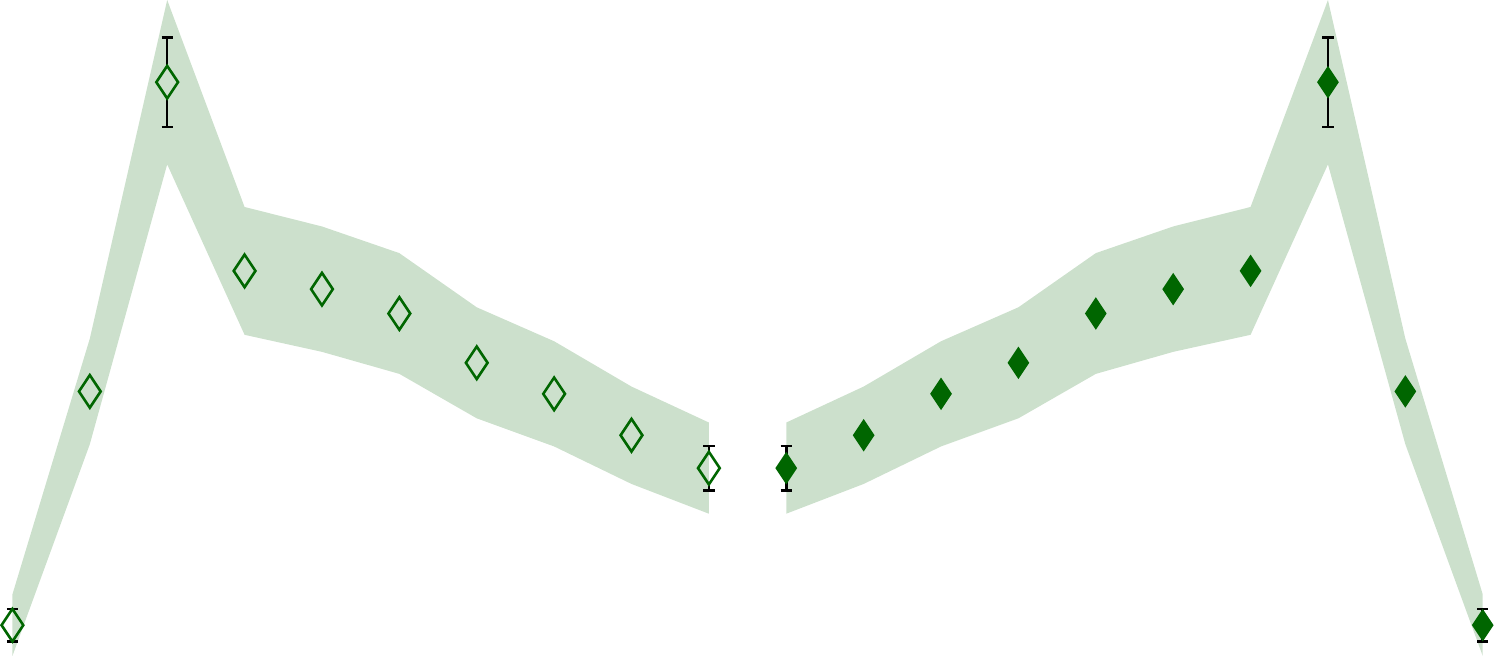}   &      \includegraphics[width=\wdth]{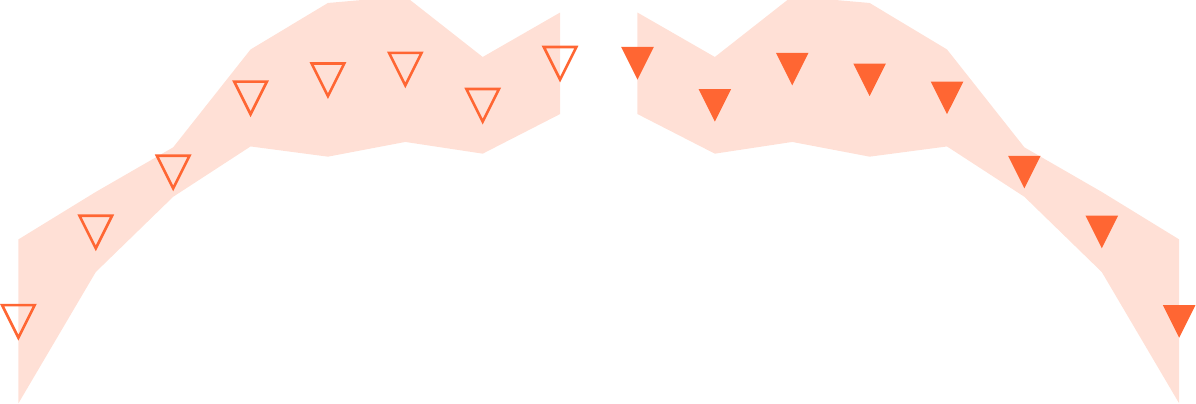}  &  \includegraphics[width=\wdth]{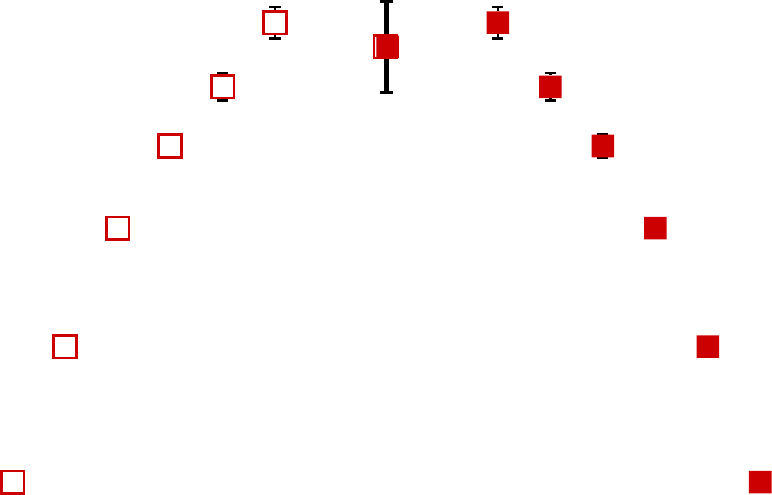}   \\ \hline
    31$A$ (30$A$)  &   7.7       &  \includegraphics[width=\wdth]{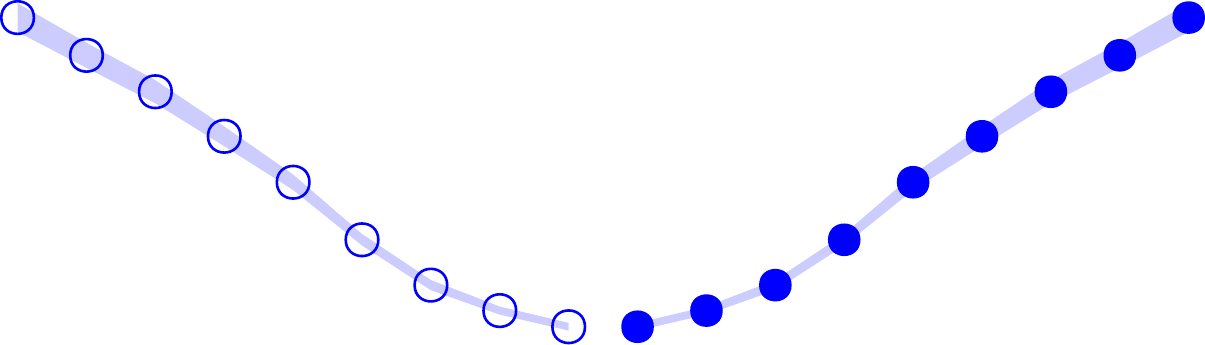} &  \includegraphics[width=\wdth]{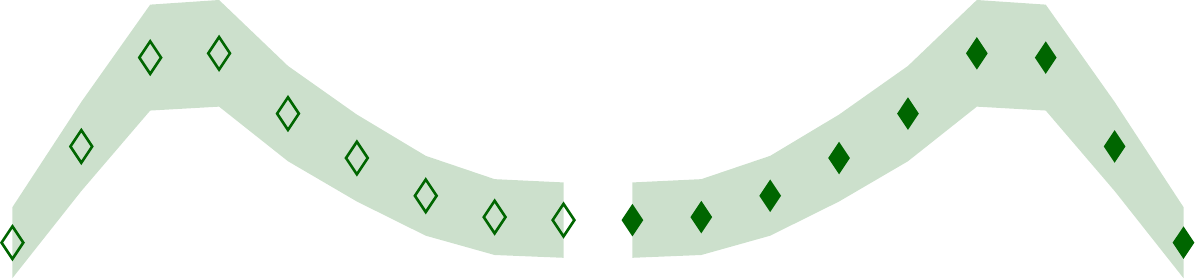}   &      \includegraphics[width=\wdth]{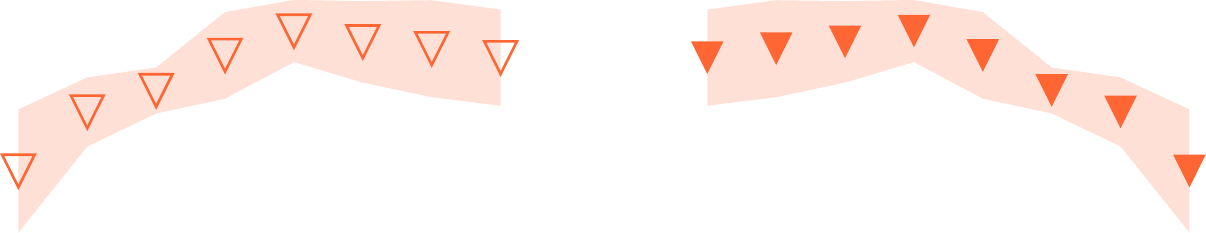}  &  \includegraphics[width=\wdth]{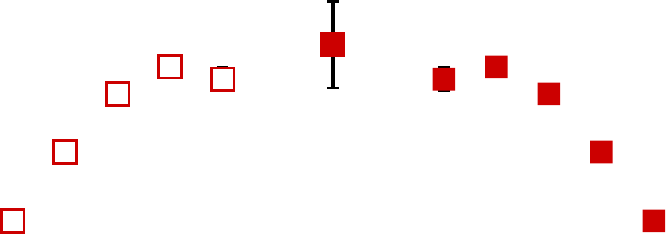}   \\ \hline
    40$A$       &  8.8       &  \includegraphics[width=\wdth]{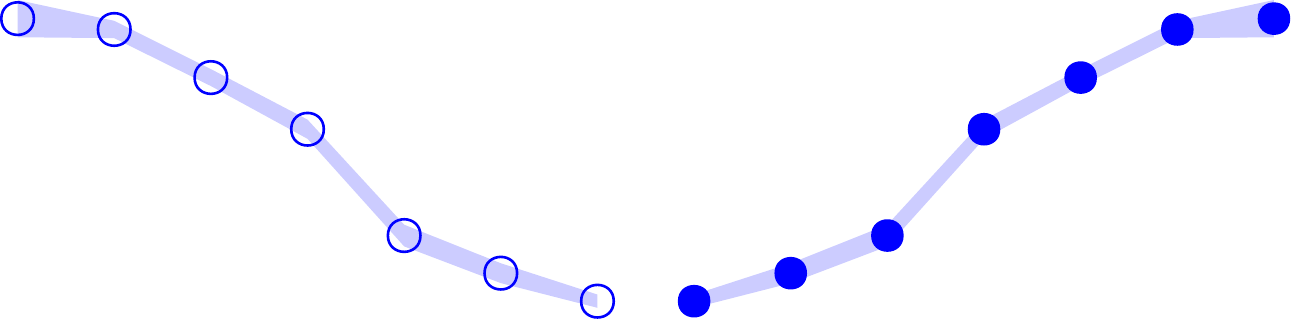} &  \includegraphics[width=\wdth]{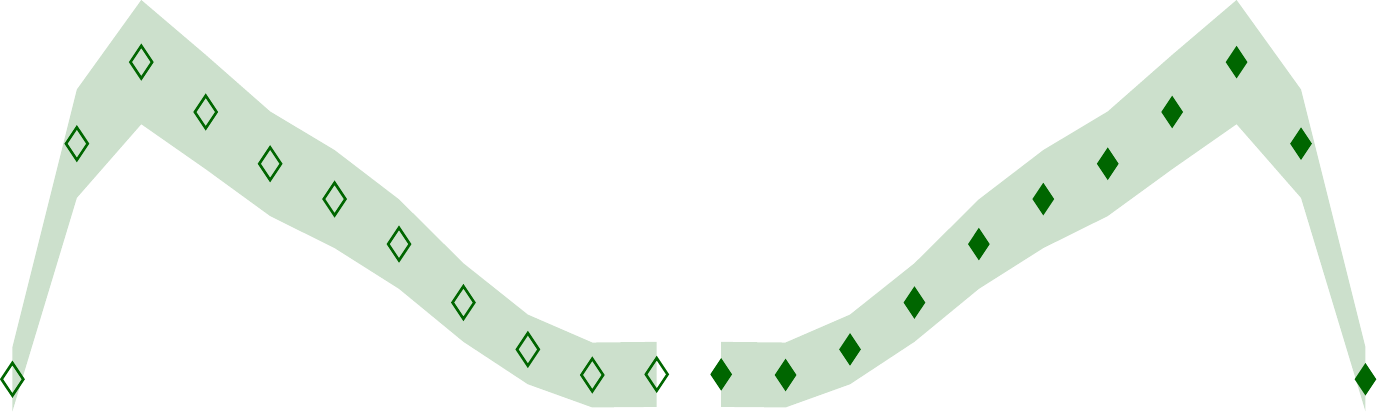}   &      \includegraphics[width=\wdth]{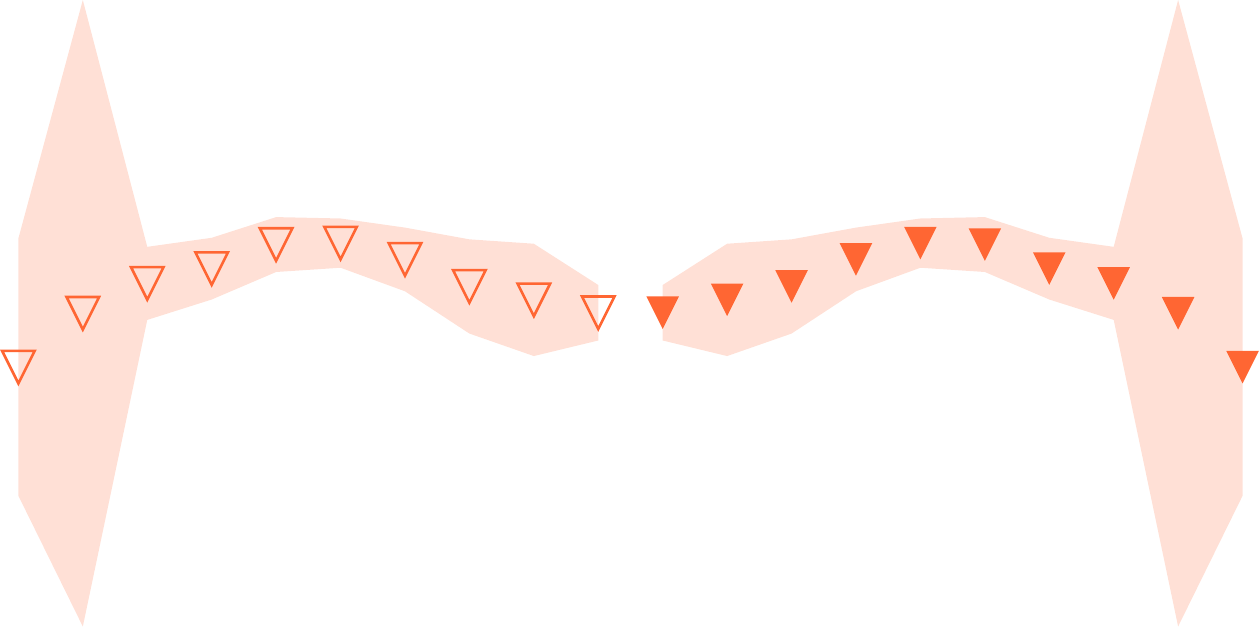}  &  \includegraphics[width=\wdth]{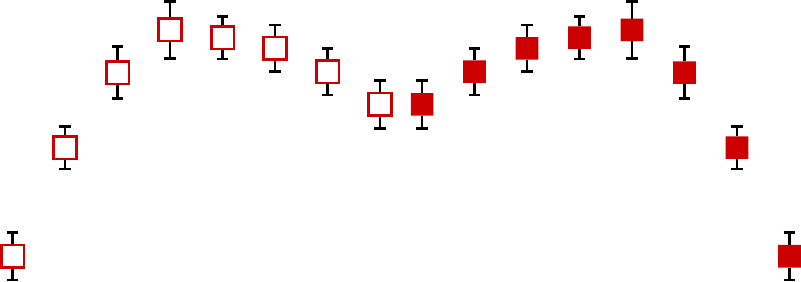}   \\ \hline
    80$A$ (75$A$) &  12.3         &  \includegraphics[width=\wdth]{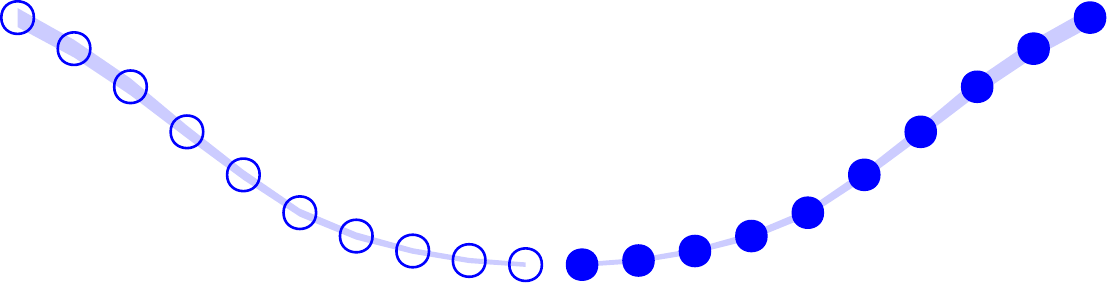} &  \includegraphics[width=\wdth]{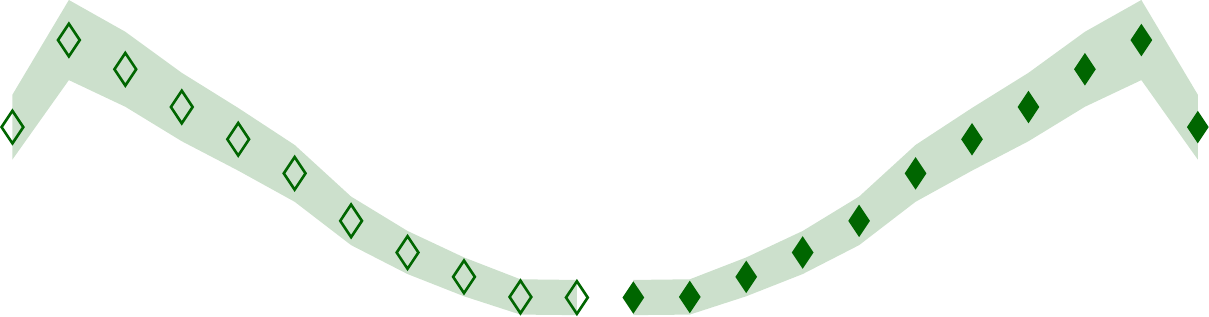}   &      \includegraphics[width=\wdth]{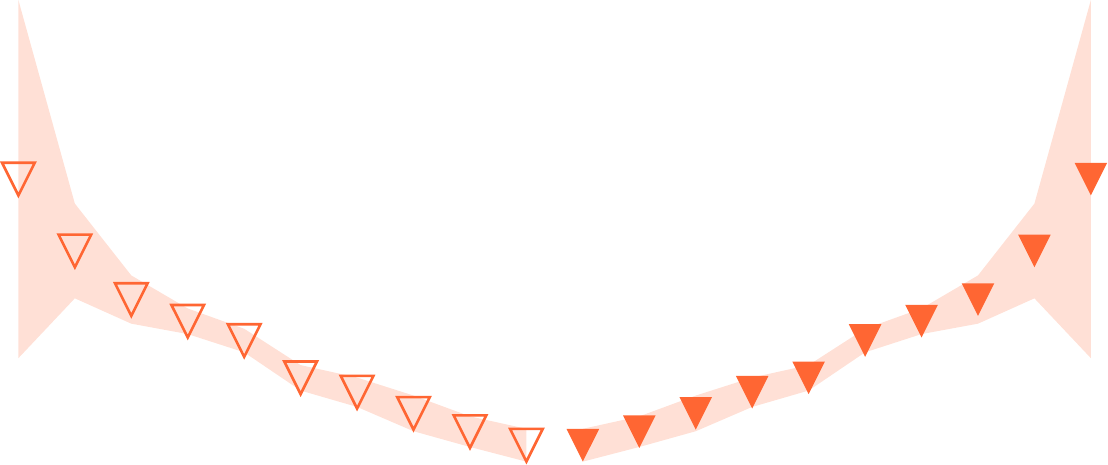}  &  \includegraphics[width=\wdth]{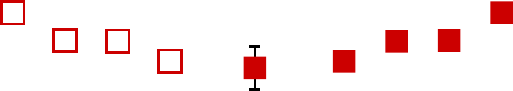}   \\ \hline
    158$A$ (150$A$)&    17.3      &  \includegraphics[width=\wdth]{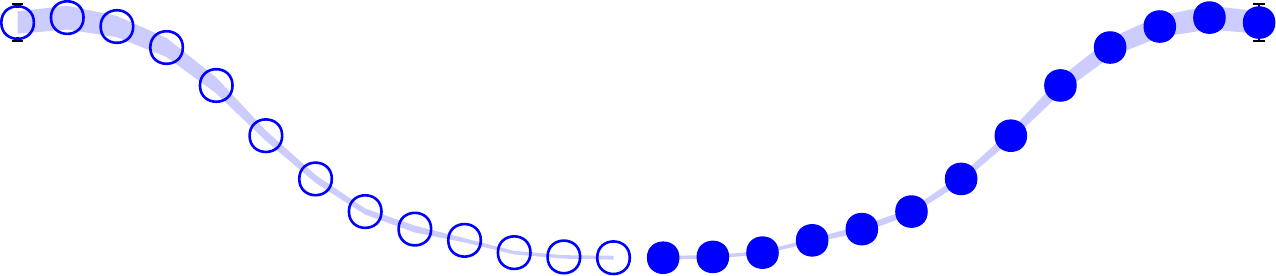} &  \includegraphics[width=\wdth]{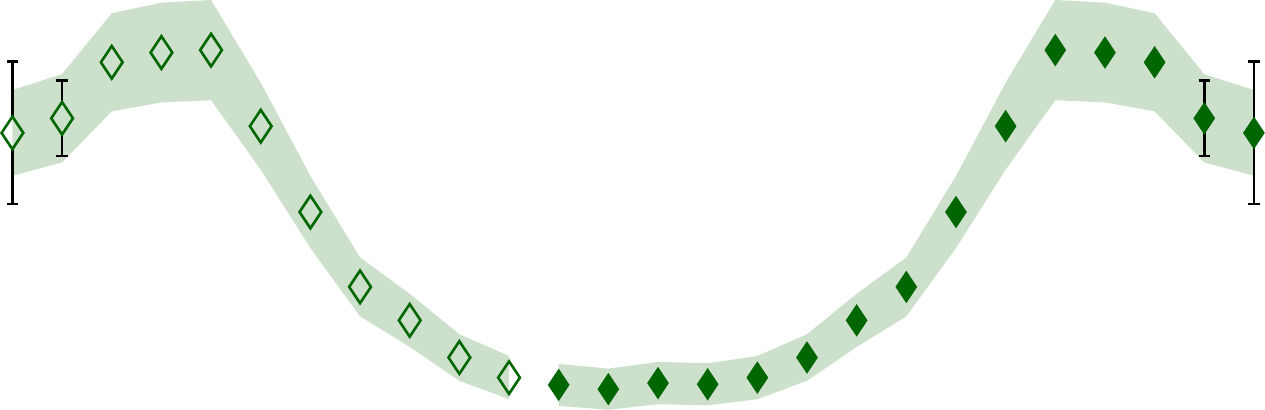}   &      \includegraphics[width=\wdth]{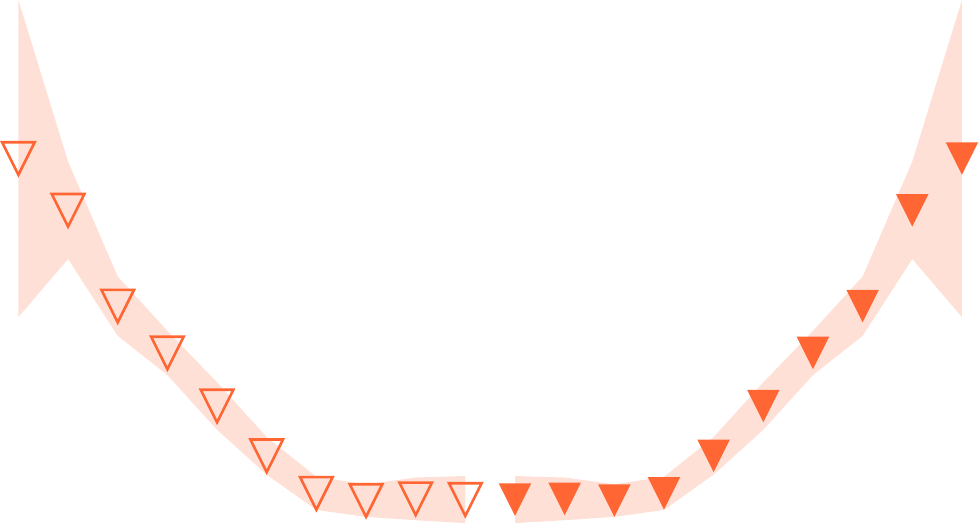}  &  \includegraphics[width=\wdth]{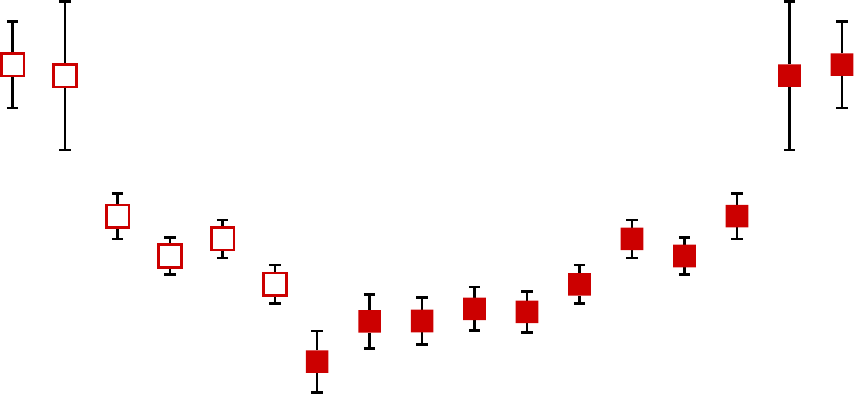}   \\ \hline
    \end{tabular}\\
    \caption{System size and energy dependence of the shape of the proton rapidity spectra.}
    \label{tab}
    \end{table}

\section{Comparison of NA61/SHINE results with models}
\label{sec:models}
\begin{figure}[h!]
        \includegraphics[width = 0.32\linewidth,trim={0cm 0cm 0cm 0cm}, clip]{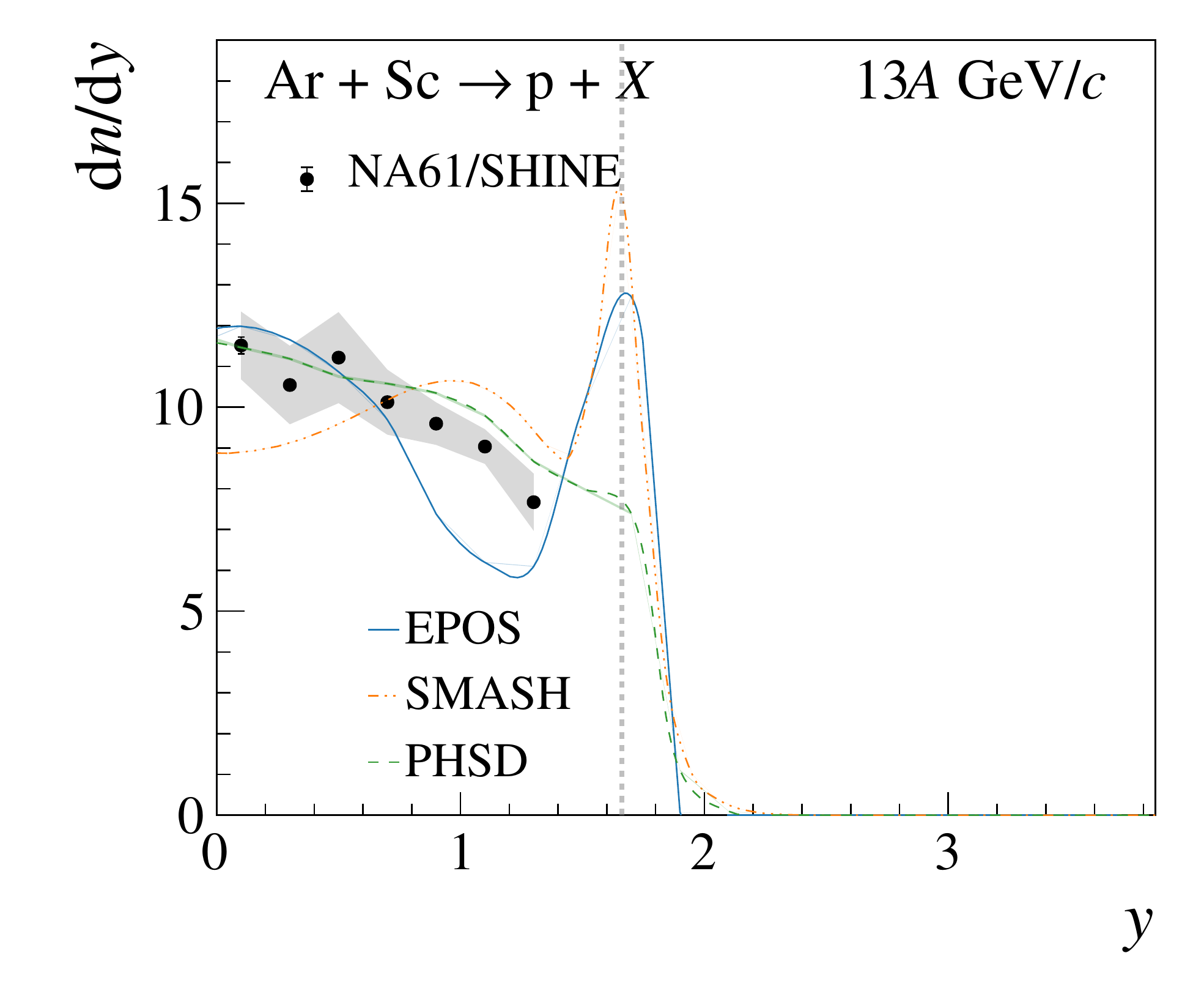}
        \includegraphics[width = 0.32\linewidth,trim={0cm 0cm 0cm 0cm}, clip]{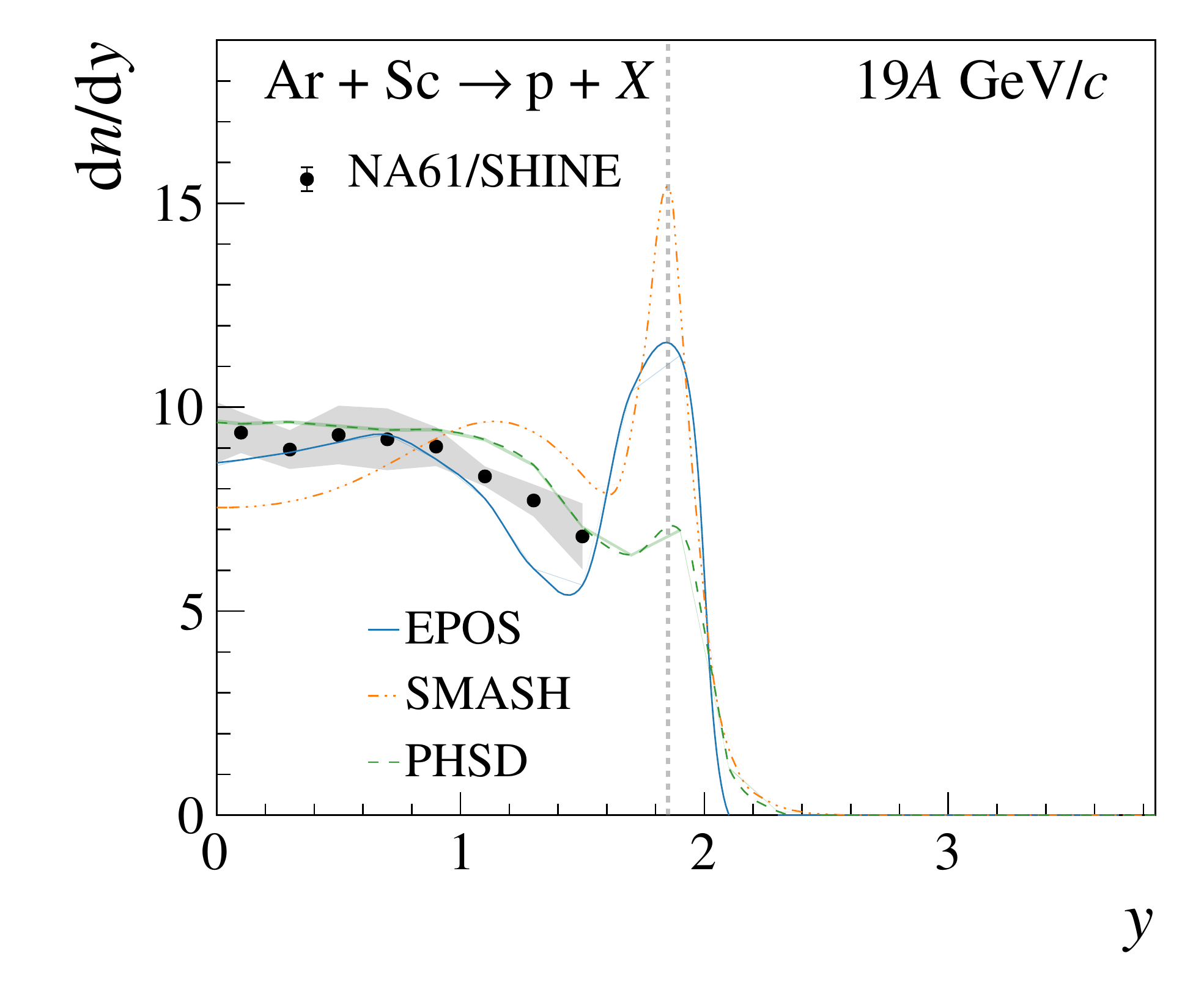}
        \includegraphics[width = 0.32\linewidth,trim={0cm 0cm 0cm 0cm}, clip]{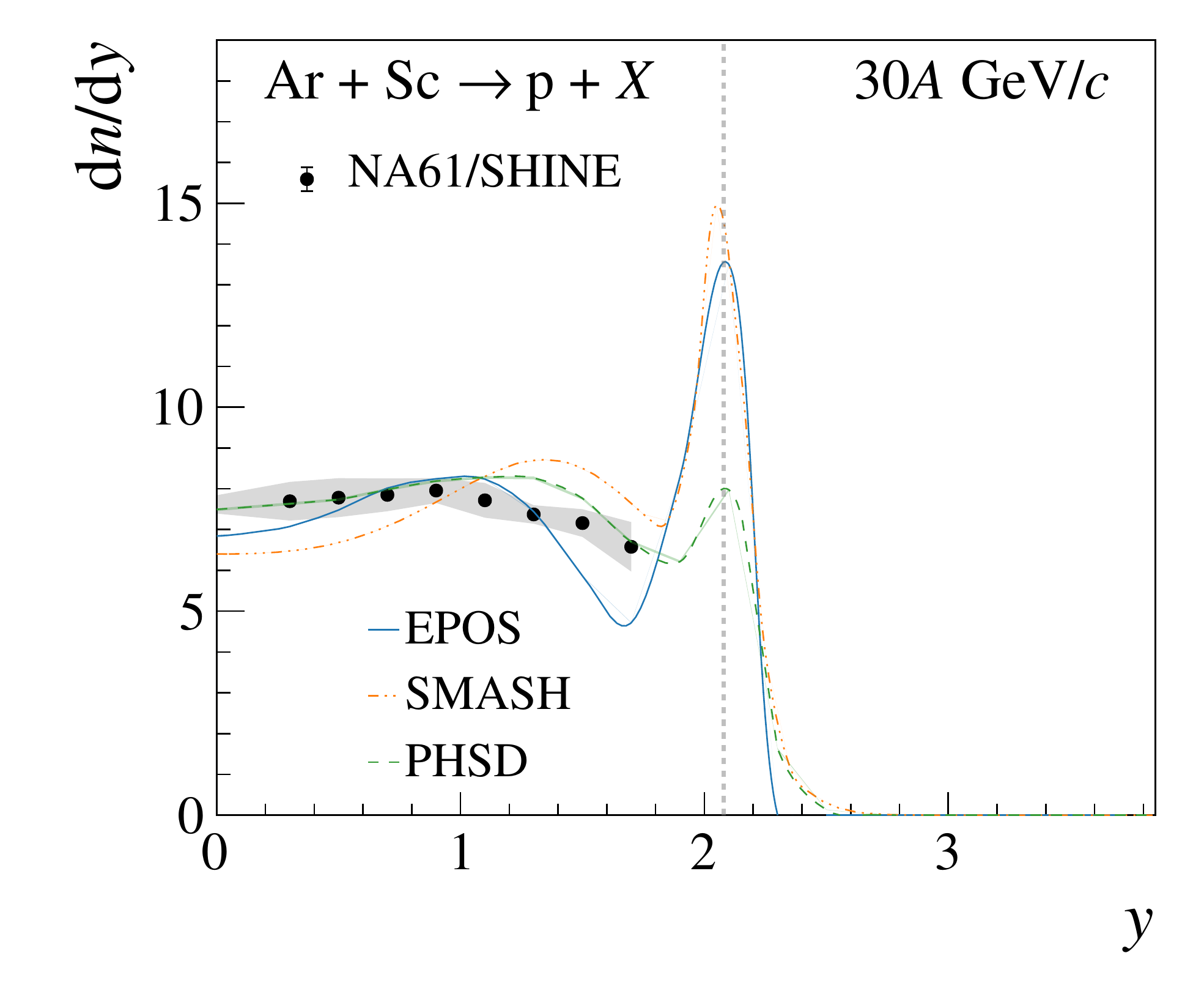}\\[-3cm]
        {\color{gray} \tiny \hspace{0.09\linewidth} preliminary \hspace{0.26\linewidth} preliminary \hspace{0.26\linewidth} preliminary}\\[25mm]
        \includegraphics[width = 0.32\linewidth,trim={0cm 0cm 0cm 0cm}, clip]{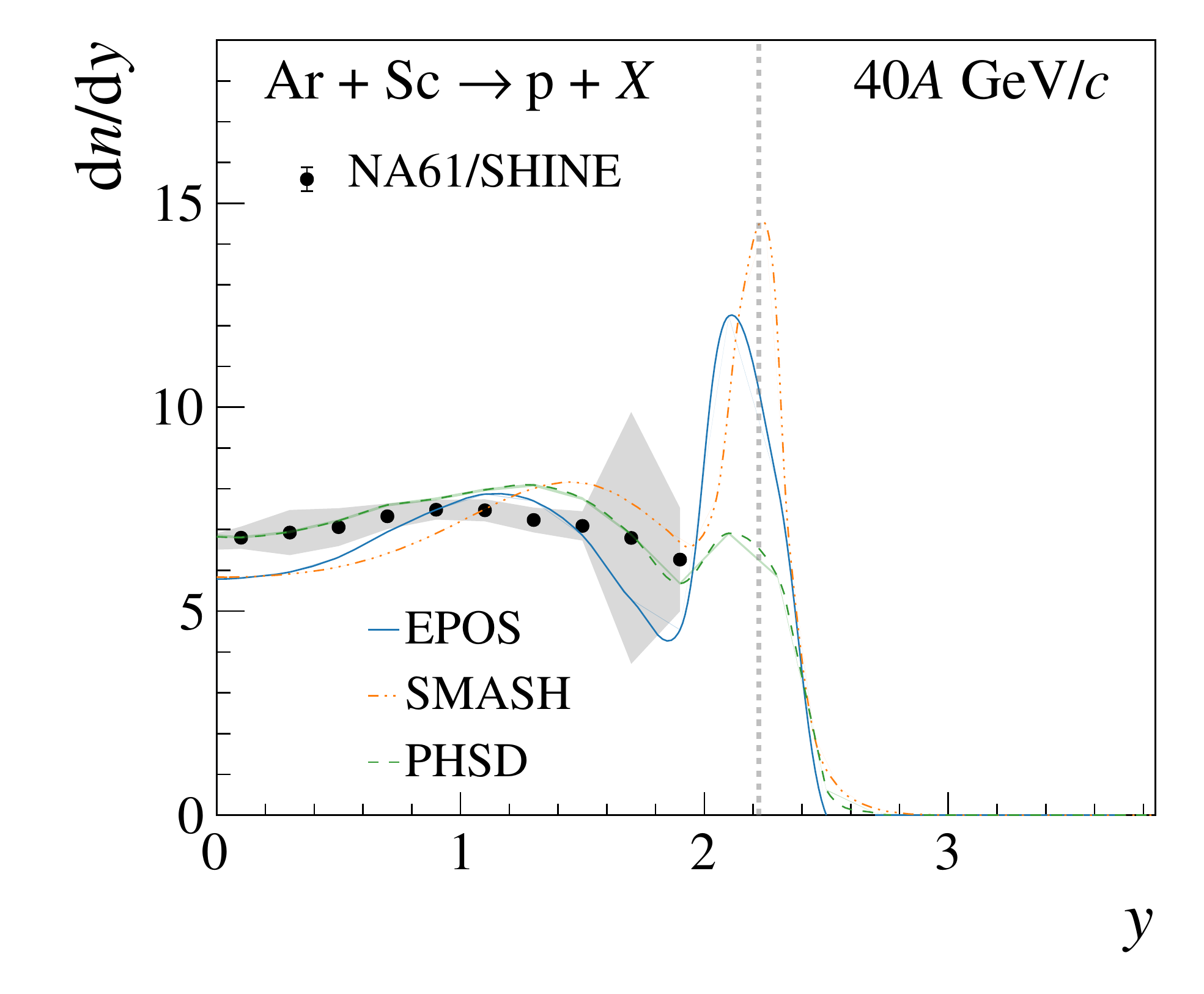}
        \includegraphics[width = 0.32\linewidth,trim={0cm 0cm 0cm 0cm}, clip]{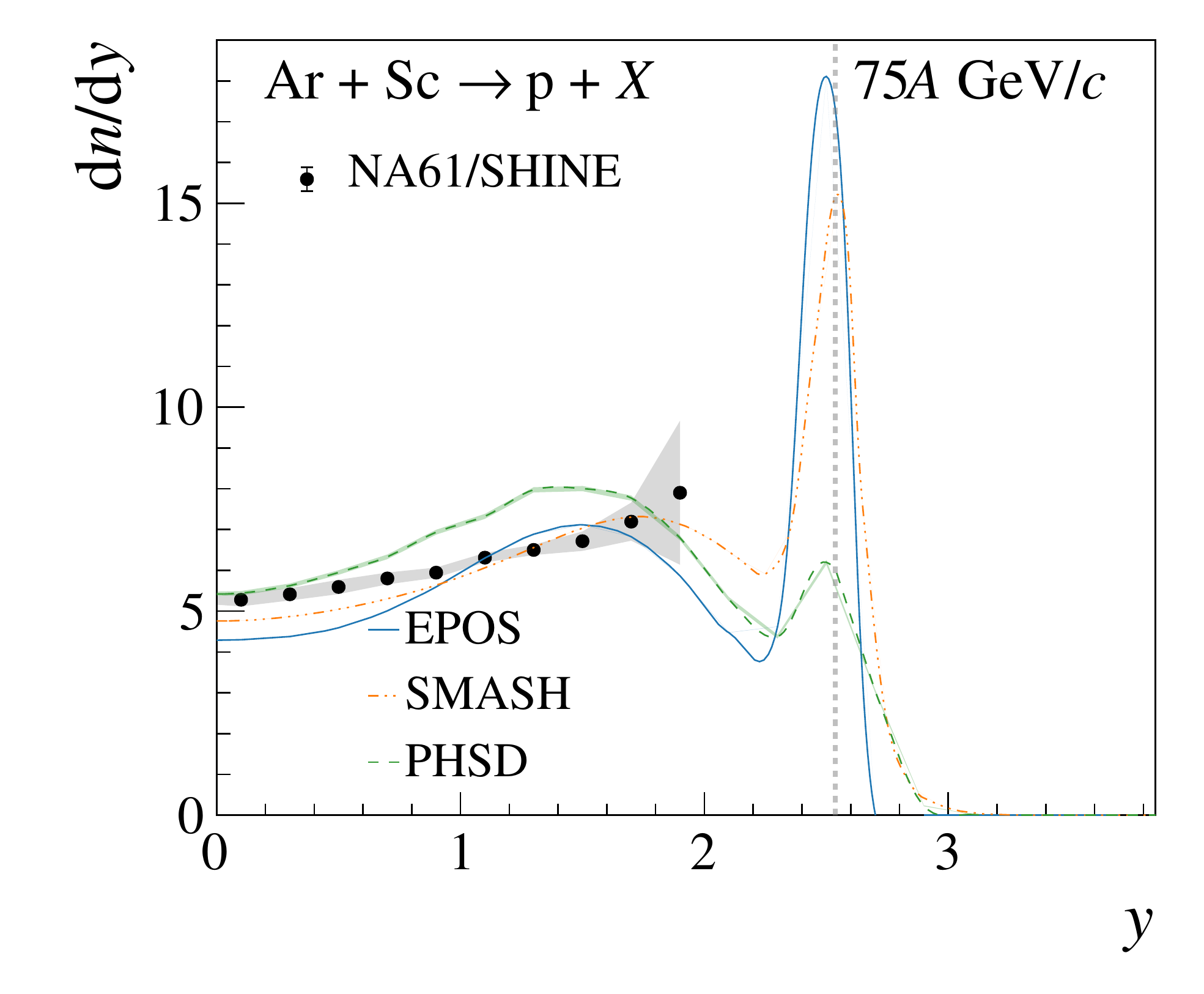}
        \includegraphics[width = 0.32\linewidth,trim={0cm 0cm 0cm 0cm}, clip]{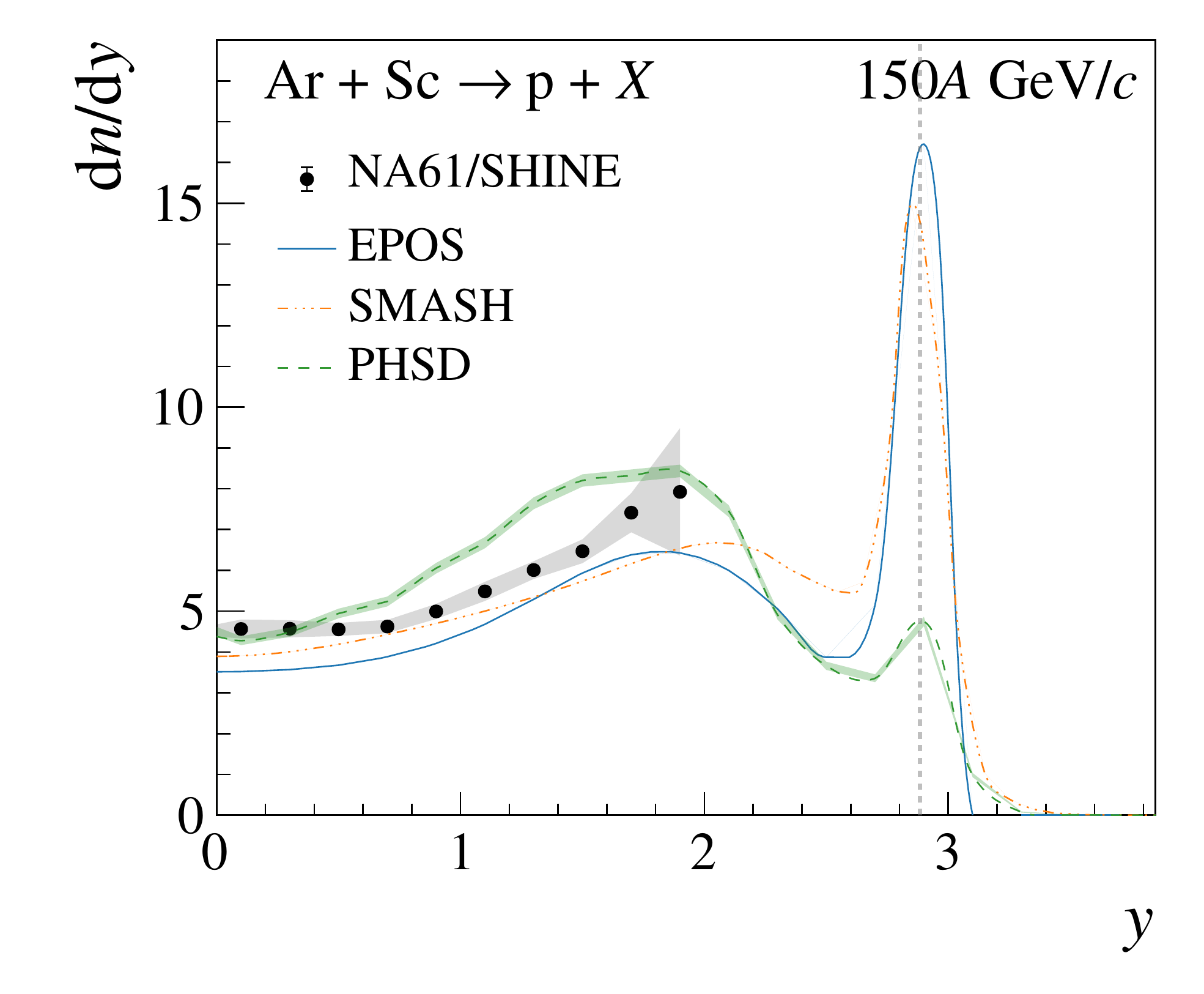}
        \\[-3cm]
        {\color{gray} \tiny \hspace{0.09\linewidth} preliminary \hspace{0.26\linewidth} preliminary \hspace{0.33\linewidth} \raisebox{2.5mm}{preliminary}}\\[20mm]
        \caption{Comparison of the proton rapidity spectra in central Ar+Sc collisions at 13$A$ -- 150$A$ GeV/$c$  with  
         \Epos 1.99 \cite{EPOS}, \PHSD 4.0 \cite{PHSD1,PHSD2} and \SmashModel 1.6 \cite{SMASH1, SMASH2} models.
        The grey dashed vertical line is plotted at the beam rapidity.}
        \label{fig:models}
    \end{figure}

The comparison of the NA61/SHINE proton rapidity spectra in \textit{central} Ar+Sc collisions with predictions of the microscopic models: \Epos 1.99 \cite{EPOS}, \PHSD 4.0 \cite{PHSD1,PHSD2} and \SmashModel 1.6 \cite{SMASH1, SMASH2} is shown in Fig. \ref{fig:models}.


The \Epos and \PHSD models describe the spectra shape relatively well, including the widely discussed in previous sections ``peak-dip'' transition. It is not the case with the \SmashModel model -- there is a large discrepancy between the data and model predictions and the distributions do not feature the peak at midrapidity at lower beam momenta. The \Epos model underestimates the yields of protons, while \PHSD describes the measured spectra shape much better, in particular at lower collision energies.

\section{Conclusions}
\label{sec:conclusions}
The dependence of the shape of the proton rapidity spectra on collision energy and the colliding system was studied. Preliminary NA61/SHINE results on the proton rapidity spectra in Ar+Sc collisions are shown for the first time.

The ``peak-dip'' transition is observed for central Ar+Sc and Pb+Pb collisions within the SPS energy range.  No such feature is observed for small systems: $p$+$p$ and Be+Be.
The ``peak-dip-peak-dip'' irregularity is observed for the heaviest systems (Pb+Pb and Au+Au), as the data is available at a wider collision energy range.
Observed features seem to uncover clear qualitative differences between the small and large systems, which are likely related to the phenomena of the onset of deconfinement and the onset of the fireball. It is important to highlight, that similar qualitative differences between small and large systems were already observed by the NA61/SHINE, in particular in the measurements of $K^+/\pi^+$ ratio -- the system of Ar+Sc is the lightest of studied systems, which shows significant enhancement of this ratio relatively to $p$+$p$ interactions \cite{arsc1,arsc2}.

More experimental data is needed to complete the studies of energy dependence for both higher and lower energies in the case of small and intermediate systems. Current studies of $^{129}$Xe+$^{139}$La by NA61/SHINE will allow to investigate the interesting region between $^{40}$Ar+$^{45}$Sc and $^{208}$Pb+$^{208}$Pb. Moreover, future SPS plans may allow performing similar studies for collisions of oxygen ions, which would provide more insight into the transition between small and large systems.

At the same time, it is paramount to study the shapes of proton rapidity spectra with modern phenomenological models, featuring a realistic equation of state, and to understand under which conditions the characteristic features of ``dips'' and ``peaks'' are to be expected.

\section*{Acknowledgements}
This work was supported by the National Science Centre, Poland (grant no. 2018/30/A/ST2/00226).

\bibliography{references.bib}

\end{document}